# Analysis of Brownian coagulation in the spatial mixing layer based on AK-iDNS framework

| | |
|---|---|
| Complete List of Authors: | Xie, Mingliang; Huazhong University of Science and Technology, State Key Laboratory of Coal Combustion |
| Abstract: | This study investigates the evolution of nanoparticle populations undergoing Brownian coagulation in a spatial mixing layer. The dynamics of particle size distribution and number concentration are analyzed using a coupled Eulerian approach that combines fluid dynamics with aerosol dynamics. The mixing layer serves as a fundamental flow configuration to understand particle-flow interactions and their effect on coagulation rates. Results demonstrate that the shear-induced spatial mixing significantly influences the spatial distribution of nanoparticles and their subsequent coagulation behavior. The enhanced mixing in the shear layer leads to locally increased particle collision frequencies, accelerating the coagulation process compared to laminar conditions. The study reveals that the evolution of the particle size distribution is strongly dependent on both the local turbulence intensity and the initial particle concentration gradients across the mixing layer. |

# Analysis of Brownian coagulation in the spatial mixing layer based on AK-iDNS framework

Xie Mingliang

State Key Laboratory of Coal Combustion, Huazhong University of Science and Technology, Wuhan 430074, China

**Abstract:** This study investigates the evolution of nanoparticle populations undergoing Brownian coagulation in a spatial mixing layer. The dynamics of particle size distribution and number concentration are analyzed using a coupled Eulerian approach that combines fluid dynamics with aerosol dynamics. The mixing layer serves as a fundamental flow configuration to understand particle-flow interactions and their effect on coagulation rates. Results demonstrate that the shear-induced spatial mixing significantly influences the spatial distribution of nanoparticles and their subsequent coagulation behavior. The enhanced mixing in the shear layer leads to locally increased particle collision frequencies, accelerating the coagulation process compared to laminar conditions. The study reveals that the evolution of the particle size distribution is strongly dependent on both the local turbulence intensity and the initial particle concentration gradients across the mixing layer.

**Keywords:** population balance equation; particle laden flow; Brownian coagulation; spatial mixing layer; average kernel method; direct numerical simulation

## Introduction

The population balance equation (PBE) is a mathematical framework utilized to describe the distribution and its dynamics of particles within a disperse system. It is extensively applied in various fields such as chemical engineering, environmental science, and biotechnology to model processes involving particulate systems, including crystallization, polymerization, aerosol dynamics, and biological cell growth (**Friedlander, 2000**). The PBE accounts for the birth, growth, aggregation, and breakage of particles, providing a comprehensive description of how the number and size distribution of particles evolve over time (**Chen et al., 2021**). By incorporating these mechanisms, the equation helps predict the behavior of complex systems where particle interactions play a crucial role. Solving the PBE can yield valuable insights into the design and optimization of industrial processes, ensuring enhanced control over product quality and process efficiency (**Shettigar et al., 2024**). However, it is computationally demanding to directly solve the PBE primarily because of its dependence on the particle volume (**Schumann, 1940**).

Over the past century, various methodologies have been developed in solving the PBE numerically to enhance understanding of particle dynamics evolution. These methodologies encompass the deterministic method of moment, sectional method and the stochastic Monte Carlo method (**Liao and Lucas, 2010**). Given that the particle number density is function of particle volume and time, the numerical methods can be categorized into two primary approaches. The first approach examines the temporal evolution of moments of particle size distribution, represented by the moment method (**Pratsinis, 1988**); the second approach investigate the particle size distribution (PSD) at extended time scales, represented by self-preserving theory (**Friedlander and Wang, 1966**). The sectional method and Monte Carlo method enable simultaneously examination of the evolution of particle size distribution with respect to time and particle volume, and are generally considered direct numerical simulation methods for investigating particle dynamics.

The sectional method divides the particle volume space into several discrete sections or bins. Within each section or bin, the size of particles can be considered as a constant or linearly distributed function, and the PBE can be solved. To achieve the required accuracy, the number of sections needs to reach hundreds or higher, which significantly increases the computational cost. Furthermore, the sectional method requires a preset range of particle size, i.e., the minimum and maximum particle volume; consequently, the volume of new particles formed after coagulation will inevitably exceed the preset maximum. Therefore, the sectional method will inevitably result in numerical divergence as the particles evolves over time. However, due to its high accuracy with a large number of sections, sectional method is often used as a reference in short-term for evaluating other methods (**Wang et al., 2007**).

The Monte Carlo method is fundamentally based on the physical model of the research object, and delineates the evolution of particles through stochastic processes and simulated events. Theoretically, it can account for the effect of finite scale, spatial correlation and fluctuations of particles. When conducting Monte Carlo simulations, it is essential to predetermine the total number of particles and their initial distribution. As particles interact, they tend to concentrate towards the central regions of the particle size distribution, while significant deviations occur at the upper and lower extremities. To enhance the accuracy of numerical results, the Monte Carlo method necessitates a substantial number of initial particles and considerable computation resources, which constitute the primarily factors limiting its widespread application in theoretical analysis and engineering practice (**Zhao and Zheng, 2011**).

The moment method transforms the particle size distribution into various moments through integral transformation, thereby converting the PBE into a set of ordinary integra-differential equations. Since the particle size distribution is equivalent to its infinite order moments, and in practical application only a limited number of moment equation can be calculated, the primary challenge for the moment method is the closure problem of the system of moment equations. To address the closure problem, researchers have developed method of moment (MOM) based on the lognormal distribution (**Lee et al., 1984**), quadrature method of moment (QMOM) (**McGraw, 1997**), Taylor series expansion method of moment (TEMOM) (**Yu et al., 2008**), and other approaches. Another significant challenge faced by the moment method is its inverse problem, i.e., how to reconstruct the particle size distribution from a set of finite order moments. Research on the inverse problem of the moment method remains an active area of investigation and a challenge in mathematics and engineering fields (**John et al., 2007**).

Despite these difficulties and challenges, the moment method is increasingly been applied to numerical simulation in particle science and technology due to its characteristics of simplicity, high efficiency and accuracy. It should be noted out that the closure of the moment equations is approached using the Taylor-series expansion technique in the TEMOM. Through constructing a system of the three first-order ordinary differential equations, the most important moments for describing the particle dynamics, including the particle number density, particle mass and geometric standard deviation, are obtained. This approach has no prior requirement for the particle size spectrum, and the limitation inherent in the lognormal distribution theory automatically eliminated. However, when dealing with certain specialized collision kernels (such as the sedimentation kernel), TEMOM cannot accurately obtain the corresponding moment models, and alternative method are required in such cases (**Pan et al., 2024**; **Xie, 2024**).

If the collision frequency function of particles is homogeneous with respect to its variables, then the PBE can be transformed into an ordinary differential equation through similarity transformation. Previous studies have revealed that the particle size distribution will reach a self-preserving form at long times (**Friedlander and Wang, 1966**). Similarity theory provides an analytical method for analyzing the asymptotic behavior of the coagulating system. Recently, Xie proposed an iterative direct numerical simulation (iDNS) to solve the governing equation of self-preserving size distribution, using the asymptotic solution of TEMOM as the initial conditions. This method establishes a one-to-one mapping relationship between TEMOM and similarity theory based on iDNS algorithm, laying a foundation for analyzing the mathematical properties and physical characteristics of the PBE (**Xie, 2023**).

To determine the size distribution of drops in fogs and clouds, an average kernel method based on binary Laplace transformation is presented by Schumann to account for the experimental results obtained (**Schumann, 1940**). Then the solution of PBE can easily be obtained analytically. The mathematical expression for the volume distribution curve shows very satisfactory agreement with the experimental values. The theory might be useful in showing how research should be directed in order to arrive at a full understanding of the mechanism by which fogs and clouds are formed. Schumann's work on solving PBE with average kernel method has become a classic in this field. So far, the result has been widely cited as the benchmarks by scholars in different fields (**Lin et al., 2018; Makoveeva and Alexandrov, 2023**). By carefully analyzing Schumann's work, it is found that the average kernel method can significantly simplifies the analytical solution of PBE. However, this method also has its own drawbacks. Firstly, the analytical solution is independent of the average kernel and the boundary condition. Secondly, there is an irreconcilable contradiction between the analytical solution and the experimental data. Although Schumann adopted the radius-based particle size distribution instead of volume-based particle size distribution to achieve the matching between analytical solution and the experimental data, this treatment did not solve the problem fundamentally. Thirdly, the binary Laplace transformation for a nonlinear collision kernel has higher computational complexity.

Coincidentally, an average kernel method coupled with iterative direct numerical simulation (AK-iDNS) framework has been proposed recently (**Pan et al., 2024)**. This method uses the analytical solution of average kernel method (AK) method as the initial condition (Xie, 2023; Pan et al., 2024), and the self-preserving size distribution is achieved through an iterative algorithm. The corrected similarity solution is not only consistent with the experimental data but also consistent with the properties of the original kernel. Therefore, the first two defects of average kernel method have been effectively addressed and overcome. Compared with the TEMOM, this new framework has the same asymptotic growth rate but in a more concise form, decoupling particle number density from other moments. Using the analytical solution of PBE with average kernel as initial condition, an iteratively corrected similarity solution satisfying the original kernel characteristics is obtained. The results not only meet the theoretical accuracy and efficiency requirements, but also agree with the experimental data. And AK- iDNS is expected to become an upgraded alternative method to TEMOM.

Research on the evolution of nanoparticles in fluid flow is essential for advancing our understanding of multiphase systems and has significant implications across various scientific and engineering disciplines. In fluid dynamics, a spatial mixing layer is a fundamental flow configuration where two parallel streams of fluid with different velocities meet and interact. This interaction creates a region of shear between the two streams, leading to the development of complex flow structures and mixing phenomenon. Compared with temporal mixing layer, the significance of studying spatial mixing layer is indeed rooted in their direct applicability to practical engineering flows (**Chorin A.J., 1968; Chorin A.J., 2000**). **Table 1** shows the key differences between temporal and spatial mixing layers across various aspects of their behavior and study. While both configurations study the same fundamental phenomenon of shear layer mixing, they offer different perspectives and are suited to different types of analysis. Temporal mixing layers are often used for fundamental studies and are easier to simulate numerically, while spatial mixing layer are more directly relevant to many practical applications and experimental studies.

The mixing layer is characterized by strong shear and coherent structures. Coherent structures refer to organized patterns of fluid motion that persist over time and space, significantly influencing the overall behavior of the flow. These structures typically manifest as vortices or eddies, which play a pivotal role in the entrainment and mixing of fluid elements. Analyzing the formation, evolution and interaction of these coherent structures provides insights into the mechanisms driving turbulence, energy transfer, and scalar mixing in various engineering and natural systems. Understanding these dynamics is essential for improving predictive models and optimizing processes in fields such as aerodynamics, environmental engineering, and industrial mixing.

Over the past few decades, significant advancements in computational methods and resources have enabled researchers to simulate temporal mixing layers with increasing accuracy and resolution (**Lele, 1992; Strang, 2007**). Early simulations focused on capturing the basic features of mixing layers, such as the growth of the shear layer and the formation of large-scale vortices. With the development of more sophisticated algorithms and an increase in computational power, it has become possible to resolve finer scales of turbulence and capture the intricate details of coherent structures within the flow. Modern numerical techniques, such as direct numerical simulation (DNS), have been particularly successful in providing high-fidelity data on the evolution of temporal mixing layers. However, performing DNS over long time periods presents several significant challenges, such as computational cost, data management, numerical stability and accuracy, physical modeling, parallelization and validation. Addressing these challenges requires advanced computational techniques, efficient algorithms, and substantial computational resources.

The integration of advanced models for particle dynamics, such as PBE, with fluid flow simulations has opened new avenues for studying the interaction between flow and particulate matter. Do the particles exhibit similar asymptotic behavior in the flow field under long-term evolution? This study aims to combine the DNS in computational fluid dynamics with AK-iDNS in particle dynamics to simulate the evolution of nanoparticles in the spatial mixing layer due to Brownian coagulation. Understanding how nanoparticles distribution and homogenize within a coherent structure can enhance the design of processes that require uniform particle distribution. In addition, another purpose of this study is to address and improve the third drawback of the average kernel method, and to provide an effective numerical method (Gauss Laguerre quadrature for double integrals) for calculating the pre-exponential factor of the average kernel (Appendix III).

## Governing equations

### Fluid fields

The flow is considered to be the constant density two-dimensional spatial mixing layer containing the nano-scale particles in the upper half as shown **Figure 1**. There are two parallel streams of fluid with different velocities ($U_1$ for the fast stream and $U_2$ for the slow stream) in the stream-wise direction ($x$ coordinate). Considering that all the perturbations vanish rapidly as $y \to \infty$, where $y$ is the coordinate in the transverse direction. Thus, the standard Fourier pseudo-spectral method can be applied directly (**Schmid and Henningson, 2001; Drain, 2002**). The primary transport variables for the flow field are the fluid velocity and pressure. These variables are governed by the Navier-Stokes equations:

$$\begin{gathered}\frac{\partial u}{\partial x}+\frac{\partial v}{\partial y}=0\\ \frac{\partial u}{\partial t}+u\frac{\partial u}{\partial x}+v\frac{\partial u}{\partial y}=-\frac{1}{\rho}\frac{\partial p}{\partial x}+\nu\left(\frac{\partial^2 u}{\partial x^2}+\frac{\partial^2 u}{\partial y^2}\right)\\ \frac{\partial v}{\partial t}+u\frac{\partial v}{\partial x}+v\frac{\partial v}{\partial y}=-\frac{1}{\rho}\frac{\partial p}{\partial y}+\nu\left(\frac{\partial^2 v}{\partial x^2}+\frac{\partial^2 v}{\partial y^2}\right)\end{gathered} \tag{1}$$

Where $\rho$ is the fluid density; $p$ is the pressure; $\nu$ is the kinematic viscosity; $u$, $v$ are the velocity component in the $x$ and $y$ directions, respectively. the initial velocity for the time developing mixing layer consists of the following two parts: the base flow $(U, V)$ and the corresponding disturbances $(u', v')$. The transverse velocity $V$ of base flow is set to zero ($V = 0$) and the streamwise velocity $U$ is specified using a hyperbolic tangent profile:

$$U=\frac{U_1+U_2}{2}+\frac{U_1-U_2}{2}\tanh\frac{y}{2\theta} \tag{2}$$

Where $\theta$ is the initial momentum thickness, and the upper and lower boundaries are slip walls. Using slip walls allows us to concentrate on the shear region without having to resolve the boundary layer on the lower and upper walls.

The governing equation of fluid flow is a second-order partial differential equation, so it requires four boundary conditions. The $u$ velocity is specified at the inlet ($x = 0$) and the outlet ($x = L_x$) boundaries. With the help of continuity equation, $\partial u/\partial x$ is also specified at the inflow and outflow boundaries. The former and the latter are known as Dirichlet and Neumann type boundary conditions, respectively. The boundary conditions are set to be zero in the transverse direction. In the numerical simulations, the instantaneous velocity component at the inlet boundary is specified using tangent hyperbolic profile which is superimposed by some perturbations (**Rogers and Moser, 1992; Moser and Rogers, 1993**). The perturbations are introduced in the form of a traveling wave. The perturbation part, which is a combination of linear eigenfunctions obtained from the linear stability calculations as shown in **Appendix I**, is specified for the inflow boundary condition. in other words:

$$u' = A * \text{Real}\,\{u'(y)\exp[i(-\omega t)]\}$$
$$v' = A * \text{Real}\,\{v'(y)\exp[i(-\omega t)]\} \tag{3}$$

Where $u'(y)$ and $v'(y)$ is the velocity eigenfunction corresponding to the most amplified mode of the two-dimensional Orr-Sommerfeld equation and $A$ is the amplitude of the two-dimensional forcing which corresponding to the fundamental frequency. Convective outflow boundary conditions are specified at the outflow. The boundary condition must be non-reflective to avoid feedback problem. The convective boundary conditions are used to generate the Dirichlet boundary conditions for both velocity components.

$$\frac{\partial u}{\partial t} = -c\frac{\partial u}{\partial x}$$
$$\frac{\partial v}{\partial t} = -c\frac{\partial v}{\partial x} \tag{4}$$

where $c$ represent the advection speed of the large-scale structures in the spatial mixing layer. The purpose of this conditions is to allow the fluid structures to flow out the domain in a natural manner. Therefore, the advection speed is chosen to match the convective velocity of the large spanwise rollers. Experimental studies of the spatial mixing layer have indicated that these large rollers advection downstream at speeds close to that of the mean speed of the layer (**Lowery and Reynolds, 1986**). Therefore, $c$ is set as $c = (U_1 + U_2)/2$ for the simulations presented here. The experimental studies also indicated that the small-scale structures move with a different speed (something close to the local mean speed of the mixing layer). For the low Reynolds number flows simulated in this work the small-scale structures are relatively large. Moreover, the results indicate that the region of the influence of the outflow boundary condition is restricted to a fairly short distance upstream of the exit plane-roughly one layer thickness. Therefore, the choice of $c$ as the mean speed of the layer is appropriate for the present study. An unforced two-dimensional mixing layer simulation whose inlet boundary contains a base profile, provided the initial conditions for the forced mixing layer simulations. A uniformly distributed Tangent hyperbolic mean velocity at all streamwise stations is the initial condition for the unforced two-dimensional mixing layer simulation. These initial and boundary conditions must be allowed to wash out before performing any statistical analysis on the layer. In other words, any particle at the inlet must be allowed to leave the outlet boundary. The mixing layer flow must also reach the statistically stationary state in which the mean velocity component is time independent.

**Particle fields**

The transport of the nano-scale particles dispersed through the fluid is governed by the particle PBE. The PBE describes the particle dynamics under the effect of different physical and chemical processes: advection, diffusion, coagulation, condensation, nucleation and other internal/external forces (**Friedlander, 2000**). In the present study, only the Brownian coagulation in the free molecule regime is considered, and the PBE can be written as:

$$\frac{\partial n}{\partial t}+u\frac{\partial n}{\partial x}+v\frac{\partial n}{\partial y}=\frac{\partial}{\partial x}\left(D_n\frac{\partial n}{\partial x}\right)+\frac{\partial}{\partial y}\left(D_n\frac{\partial n}{\partial y}\right)+\left[\frac{\partial n}{\partial t}\right]_{coag} \tag{5}$$

where the source term $[\partial n/\partial t]_{coag}$ is given by the classical Smoluchowski equation as

$$\left[\frac{\partial n}{\partial t}\right]_{coag}=\frac{1}{2}\int_0^v \beta(v_1,v-v_1)n(v_1)n(v-v_1)dv_1-\int_0^\infty \beta(v_1,v)n(v_1)n(v)dv_1 \tag{6}$$

which represents the effects of particle-particle interactions resulting in coagulation, and $n(v,t)dv$ is the number density of particles per unit spatial volume with particle volume from $v$ to $v+dv$ at time $t$; and $\beta$ is the collision kernel of coagulation. For brownian coagulation in the free molecule regime, the collision kernel takes:

$$\beta=\left(\frac{3}{4\pi}\right)^{\frac{1}{6}}\left(\frac{6k_BT}{\rho_p}\right)^{\frac{1}{2}}\left(\frac{1}{v_i}+\frac{1}{v_j}\right)^{\frac{1}{2}}\left(v_i^{\frac{1}{3}}+v_j^{\frac{1}{3}}\right)^2 \tag{7}$$

where $k_B$ is the Boltzmann's constant; $T$ is the fluid temperature; $\rho_p$ is the particle density; and $D_n$ is the particle diffusion coefficient, which is given by Einstein-Smoluchowski relation as

$$D_n=\frac{k_BT}{f} \tag{8}$$

where $f$ is the friction coefficient of particles in fluids. In the free molecule regime, the friction coefficient can be derived from the kinetic theory as (**Epstein, 1924**)

$$f=\frac{2}{3}d_p^2\rho\left(\frac{2\pi k_BT}{m}\right)^{\frac{1}{2}}\left(1+\frac{\pi\alpha_p}{8}\right) \tag{9}$$

where $m$ is the molecular mass of fluid molecules, $\rho$ is the density of fluid, $\alpha_p$ is the accommodation coefficient, and $d_p$ is the diameter of particles, and $d_p=\sqrt[1/3]{6v/\pi}$.

## AK-iDNS framework for PBE

### The average kernel method (AK)

To solve the Smoluchowski equation with collision kernel depending on the particle size, Schumann proposed the average kernel method (**Schumann, 1940**), which is defined with binary Laplace transformation as

$$\int_0^\infty\int_0^\infty \bar{\beta}\exp\left(-\frac{v_1+v}{v_a}\right)dv_1dv=\int_0^\infty\int_0^v \beta(v_1,v)\exp\left(-\frac{v_1+v}{v_a}\right)dv_1dv \tag{10}$$

where $v_a$ is the algebraic mean volume of particle size distribution. Through operation and reorganization, the average kernel can be represented as

$$\bar{\beta}=\frac{1}{v_a^2}\int_0^\infty\int_0^v \beta(v_1,v)\exp\left(-\frac{v_1+v}{v_a}\right)dv_1dv \tag{11}$$

For homogeneous collision kernel, it has the following properties:

$$\begin{cases} \beta(\alpha v, \alpha v_1) = \alpha^q \beta(v, v_1); \\ \beta(v, v_1) = \beta(v_1, v); \\ \beta(v, v_1) \geq 0; \end{cases} \tag{12}$$

in which $q$ is power index, and $\alpha$ is the scale factor. It is easy to prove that the homogeneous collision kernels satisfy the following differential equation:

$$v\frac{\partial \beta}{\partial v} + v_1\frac{\partial \beta}{\partial v_1} - q\beta = 0 \tag{13}$$

If the scale factor is $\alpha = 1/v_a$, the collision kernel can be expressed as

$$\beta(v, v_1) = v_a^q\, \beta\left(\frac{v}{v_a}, \frac{v_1}{v_a}\right) \tag{14}$$

Let the dimensionless particle volume as

$$\eta = \frac{v}{v_a} \tag{15}$$

And the average kernel can be expressed as

$$\bar{\beta} = p v_a^q \tag{16}$$

where $p$ is a proportional factor, and it can be calculated as

$$p = \int_0^\infty \int_0^{v_a \eta} e^{-\eta-\eta_1}\, \beta(\eta, \eta_1) d\eta_1 d\eta \tag{17}$$

which usually represents the total collision frequency of particle coagulating system. Due to the symmetry of the homogeneous collision kernel, it can be simplified as

$$p = \frac{1}{2}\int_0^\infty \int_0^\infty e^{-\eta-\eta_1}\, \beta(\eta, \eta_1) d\eta_1 d\eta \tag{18}$$

For Brownian coagulation in the free molecule regime, the parameters $p$ and $q$ can be calculated as (**Pan et al., 2024**)

$$p = 4\sqrt{2}B_1\Gamma\left(\frac{13}{12}\right), \qquad q = 1/6 \tag{19}$$

where $\Gamma$ is the Euler gamma function, and the average kernel can be expressed as $\bar{\beta} = 4\sqrt{2}B_1\Gamma\left(\frac{13}{12}\right) v_a^{\frac{1}{6}}$.

**The moment model based on AK**

In the moment method, the $k$th order moment of the volume-based PSD is defined as

$$M_k(t) = \int_0^\infty v^k n(v, t) dv \tag{20}$$

Using the moment transformation above, the Smoluchowski equation is transformed into a series of ordinary differential equations as

$$\frac{dM_k}{dt} = -\frac{1}{2}\int_0^\infty \int_0^\infty \left[(v + v_1)^k - v^k - v_1^k\right] \beta(v, v_1) n(v) n(v_1) dv dv_1 \tag{21}$$

Generally, the minimum set of moments required to close the moment equations is the first three, i.e., $M_0$, $M_1$ and $M_2$. The zeroth order moment ($M_0$) represents the total particle number concentration; the first moment ($M_1$) is proportional to the particle mass concentration; and the second moment ($M_2$) describes the dispersity of PSD.

Substituting the average kernel $\bar{\beta} = p{v_a}^q$ into the moment equations, it can be found

$$\begin{cases} \left[\frac{dM_0}{dt}\right]_{coag} = -\frac{1}{2}\bar{\beta}M_0^2 \\ \left[\frac{dM_1}{dt}\right]_{coag} = 0. \\ \left[\frac{dM_2}{dt}\right]_{coag} = \bar{\beta}M_1^2 \end{cases} \tag{22}$$

Usually, $v_a$ is selected as the algebraic mean volume of PSD, its expression can be written as

$$v_a = \frac{M_1}{M_0} \tag{23}$$

For the classical Smoluchowski equation, $M_1$ remains constant due to the rigorous mass conservation requirement, and the moment equations is decoupled. And the evolution of particle number density can be obtained with only one equation. According to the value of index $q$, the solution of zeroth order moment can be divided into three types. In the present study, the index $q = 1/6$, then analytical solution can be found as

$$M_0 = \left[M_{00}^{q-1} - \frac{q-1}{2}pM_1^q(t - t_0)\right]^{\frac{1}{q-1}} \tag{24}$$

The corresponding asymptotic scaling growth rate of the $M_0$ can be found as

$$\frac{1}{M_0}\frac{dM_0}{dt} = \frac{1}{(q-1)t} \tag{25}$$

**Similarity transformation and corrected similarity solution**

The similarity transformation is based on the assumption that the fraction of the particles in a given size range is only volume dependent on the dimensionless particle volume ($\eta = v/v_a$), then the similarity transformation takes the form as (**Friedlander and Wang, 1966**)

$$\begin{cases} \eta = \frac{v}{v_a} \\ \psi(\eta) = \frac{M_1}{M_0^2}n(v,t) \end{cases} \tag{26}$$

in which $\eta$ is dimensionless particle volume, $\psi$ is the dimensionless particle size distribution, and the Smoluchowski equation is transformed into the following governing equation of self-preserving theory (**Xie, 2023**):

$$A\left[2\psi(\eta) + \eta\frac{d\psi(\eta)}{d\eta}\right] = C(\eta) - \psi(\eta)g(\eta) \tag{27}$$

in which the total collision frequency $A$, the gain term $C(\eta)$, and the loss term $g(\eta)$ are given as follows:

$$\begin{cases} A = -\frac{1}{2}\int_0^\infty\int_0^\infty \beta(u\eta, u\eta_1)\,\psi(\eta)\psi(\eta_1)d\eta d\eta_1 \\ C(\eta) = \frac{1}{2}\int_0^{u\eta} \beta\big(u\eta_1, u(\eta - \eta_1)\big)\psi(\eta_1)\psi(\eta - \eta_1)d\eta_1 \\ g(\eta) = \int_0^\infty \beta(u\eta_1, u\eta)\psi(\eta_1)d\eta_1 \end{cases} \tag{28}$$

The dimensionless distribution function $\psi(\eta)$ to be solved is restricted by the following mathematical and physical constraints:

$$\int_0^\infty \psi(\eta)d\eta = 1\,; \int_0^\infty \eta\psi(\eta)d\eta = 1\,; \psi(\eta) \geq 0 \tag{29}$$

and the boundary conditions:

$$\begin{cases} \psi(\eta) \to 0 \ for \ \eta \to \infty \\ \eta\psi(\eta) \to 0 \ for \ \eta \to 0 \end{cases} \tag{30}$$

The existence of the similarity solution is related to the asymptotic properties of the kernel. Substituting the average kernel $\bar{\beta} = pu^q$ into the governing equation, it can be found that

$$\begin{cases} A = -\frac{1}{2}pu^q \\ C(\eta) = \frac{1}{2}pu^q \int_0^{u\eta} \psi(\eta_1)\psi(\eta - \eta_1)d\eta_1 \\ g(\eta) = pu^q \end{cases} \tag{31}$$

Therefore, the governing equation can be simplified as

$$\eta \frac{d\psi(\eta)}{d\eta} + \int_0^{\eta} \psi(\eta_1)\psi(\eta - \eta_1)d\eta_1 = 0 \tag{32}$$

Using the Laplace transformation:

$$\begin{cases} \mathcal{L}[\psi(\eta)] = \int_0^{\infty} \psi(\eta)\, e^{-s\eta} d\eta = \Psi(s) \\ \mathcal{L}\left[\eta \frac{d\psi(\eta)}{d\eta}\right] = -s\frac{d\Psi(s)}{ds} - \Psi(s) \\ \mathcal{L}\left[\frac{d\psi(\eta)}{d\eta}\right] = -s\Psi(s) - \psi(0) \\ \mathcal{L}\left[\int_0^{\eta} \psi(\eta_1)\psi(\eta - \eta_1)d\eta_1\right] = \Psi(s)^2 \end{cases} \tag{33}$$

where $\mathcal{L}$ is the Laplace operator, $s$ is the Laplace transform parameter, $\Psi$ is the transformed dimensionless PSD, and $\psi(0) = 0$ for realistic kernel. Then the transformed governing equation of self-preserving size distribution can be rewritten as

$$-s\frac{d\Psi(s)}{ds} - \Psi(s) + \Psi(s)^2 = 0 \tag{34}$$

Its solution can be found as

$$\Psi(s) = \frac{1}{1+s} \tag{35}$$

Its inverse Laplace transformation is defined as

$$\mathcal{L}^{-1}[\Psi(s)] = \int_0^{\infty} \Psi(s)\, e^{s\eta} ds = \psi(\eta) \tag{36}$$

For classical Smoluchowski equation, the similarity solution can be found as

$$\psi(\eta) = e^{-\eta} \tag{37}$$

Based on the inverse similarity transformation, the solution of classical Smoluchowski equation can be written as

$$n(v,t) = \frac{M_0^2}{M_1} \exp\left[-\frac{M_0}{M_1}v\right] \tag{38}$$

Usually, the analytical similarity solution is independent of kernels and the boundary condition. therefore, the obtained solution may not match the experimental data or carry no actual physical significance. An improved method is to obtain the corrected similarity solution by iDNS using the present analytical similarity solution as initial condition, the physically realistic kernel and the corresponding boundary condition can be incorporated. And the simplified AK-iDNS algorithm is listed as below.

| Algorithm: the AK-iDNS framework for SCE with initial distribution $\psi(\eta) = e^{-\eta}$ |
|---|
| input: |
| maximum value of $\eta$ |
| numerical step $\Delta\eta$ |
| coagulation kernel $\beta$ |
| preset error and its limit $error_1, error_2, \varepsilon$ |
| initial $\psi(\eta) = e^{-\eta}, A, C(\eta), g(\eta)$ |
| output: |
| while $error_2 > \varepsilon$ |
| while $error_1 > \varepsilon$ |
| while $\psi(\eta) > 0$ |
| updating $A, g(\eta), C(\eta), \psi(\eta)$ |
| updating $\int_0^\infty \psi(\eta)d\eta\,; \int_0^\infty \eta\psi(\eta)d\eta$ |
| calculating $error_1 = \max\left(\left\lvert\int_0^\infty \psi(\eta)d\eta - 1\right\rvert, \left\lvert\int_0^\infty \eta\psi(\eta)d\eta - 1\right\rvert\right)$ |
| calculating $error_2 = \lvert updated\ A - initial\ A\rvert$ |
| end while |
| end while |
| end while |

A detailed description of AK-iDNS for the classical Smoluchowski coagulation equation can be found in our previous work (**Xie, 2023; Pan et al., 2024; Xie, 2024**).

## Diffusion coefficient based on particle moment

The diffusion coefficient of particles in a fluid is a measure of how quickly particles spread out due to random motion. For nanoparticles, this can be influenced by their size, which can be characterized by the moments of the particle size distribution.

For spherical particles in the free molecule regime, the diffusion coefficient $D_n$ can be estimated using Einstein-Smoluchowski relation $D_n = k_B T/f$ . To relate the diffusion coefficient to the moments of the particle size distribution, we can use an average volume or diameter. One common choice is the volume-weighted mean diffusion coefficient as

$$\overline{D_n} = \int_0^\infty D_n \frac{n(v,t)}{M_0} dv = \frac{1}{\frac{2}{3}\left(\frac{6}{\pi}\right)^{\frac{2}{3}} \rho \left(\frac{2\pi k_B T}{m}\right)^{\frac{1}{2}} \left(1+\frac{\pi\alpha_p}{8}\right)} \frac{1}{M_0} M_{-\frac{2}{3}} \tag{39}$$

The fractional particle moment can be expressed as (**Xie, 2016**)

$$M_k = \frac{M_1^k}{M_0^{k-1}}\left[1 + \frac{k(k-1)(M_C-1)}{2}\right] \tag{40}$$

where the dimensionless particle moment ($M_C$) is defined as

$$M_C = \frac{M_0 M_2}{M_1^2} \tag{41}$$

and

$$M_{-\frac{2}{3}} = \frac{4+5M_C}{9} M_0 v_a^{-\frac{2}{3}} \tag{42}$$

Then the average diffusion coefficient is

$$\overline{D_n} = \frac{1}{\frac{2}{3}\left(\frac{6}{\pi}\right)^{\frac{2}{3}}\rho\left(\frac{2\pi k_B T}{m}\right)^{\frac{1}{2}}\left(1+\frac{\pi\alpha p}{8}\right)}\frac{4+5M_C}{9}v_a^{-\frac{2}{3}} \tag{43}$$

This relationship highlights how the diffusion coefficient depends on the statistical properties of the particle size distribution, providing a link between microscopic particle characteristics and macroscopic transport properties.

**Non-dimensionalization**

The governing equations are non-dimensionalized to simplify the treatment and analysis of the interaction between fluid and particle fields. It can be accomplished using the following relations:

$$t^* = \frac{t}{L/U}; x^* = \frac{x}{L}; y^* = \frac{y}{L}; u^* = \frac{u}{U}; v^* = \frac{v}{U}; p^* = \frac{p}{\rho U^2}; M_k^* = \frac{M_k}{M_{k0}}; v_a^* = \frac{v_a}{v_{a0}}; \tag{44}$$

In which the characteristic length $L$ is the initial momentum thickness of the mixing layer ($\theta$); the characteristic velocity $U$ is the velocity difference across the mixing layer ($(U_1 + U_2)/2$); $M_{k0}$ is the initial value of the *k*th moment. Substituting the relations given in above into the governing equations yields the familiar mass and momentum conservation equations (For brevity, the star symbol '*' is omitted thereafter):

$$\begin{gathered}\frac{\partial u}{\partial x} + \frac{\partial v}{\partial y} = 0\\ \frac{\partial u}{\partial t} + u\frac{\partial u}{\partial x} + v\frac{\partial u}{\partial y} = -\frac{\partial p}{\partial x} + \frac{1}{Re}\left(\frac{\partial^2 u}{\partial x^2} + \frac{\partial^2 u}{\partial y^2}\right)\\ \frac{\partial v}{\partial t} + u\frac{\partial v}{\partial x} + v\frac{\partial v}{\partial y} = -\frac{\partial p}{\partial y} + \frac{1}{Re}\left(\frac{\partial^2 v}{\partial x^2} + \frac{\partial^2 v}{\partial y^2}\right)\end{gathered} \tag{45}$$

Where the Reynolds number is $Re = UL/\nu$. Similarly, the non-dimensionalized equations for the first three moment equations of the particle fields are given by (**Garrick et al., 2006; Settumba and Garrick, 2003; Xie et al., 2012; Tsagkaridis et al., 2022**)

$$\begin{cases}\frac{\partial M_0}{\partial t} + u\frac{\partial M_0}{\partial x} + v\frac{\partial M_0}{\partial y} = \frac{1}{ReSc}\frac{\partial}{\partial x}\left(v_a^{-\frac{2}{3}}\frac{\partial M_0}{\partial x}\right) + \frac{1}{ReSc}\frac{\partial}{\partial y}\left(v_a^{-\frac{2}{3}}\frac{\partial M_0}{\partial y}\right) - \frac{1}{2}Da v_a^{\frac{1}{6}} M_0^2\\ \frac{\partial M_1}{\partial t} + u\frac{\partial M_1}{\partial x} + v\frac{\partial M_1}{\partial y} = \frac{1}{ReSc}\frac{\partial}{\partial x}\left(v_a^{-\frac{2}{3}}\frac{\partial M_1}{\partial x}\right) + \frac{1}{ReSc}\frac{\partial}{\partial y}\left(v_a^{-\frac{2}{3}}\frac{\partial M_1}{\partial y}\right)\\ \frac{\partial M_2}{\partial t} + u\frac{\partial M_2}{\partial x} + v\frac{\partial M_2}{\partial y} = \frac{1}{ReSc}\frac{\partial}{\partial x}\left(v_a^{-\frac{2}{3}}\frac{\partial M_2}{\partial x}\right) + \frac{1}{ReSc}\frac{\partial}{\partial y}\left(v_a^{-\frac{2}{3}}\frac{\partial M_2}{\partial y}\right) + Da v_a^{\frac{1}{6}} M_1^2\end{cases} \tag{46}$$

And the Schmidt number based on the particle moment is given as

$$Sc = \frac{\nu}{\kappa} v_{a0}^{\frac{2}{3}} \tag{47}$$

And the size independent diffusivity ($\kappa$) is

$$\kappa = \overline{D_n} v_a^{\frac{2}{3}} \tag{48}$$

The Damkohler number ($Da$) represents the ratio of the convective time scale to the coagulation time scale and is given by

$$Da = \frac{p M_{00}^2 v_{a0}^{\frac{1}{6}}}{U/L} \tag{49}$$

It is obvious that equations (45) are the system of partial difference equations and all the terms are denoted by the first three moments $M_0$, $M_1$ and $M_2$, and the system presents non closure problem. Under these conditions, the first three moments for describing particle dynamics are obtained through solving the systems of partial differential equations. Here, the deviation of equations (45) for particle fields does not involve any assumptions for the particle size distribution, and the final mathematical form is much simpler than the method of moment, QMOM, TEMOM, etc. Furthermore, only the first and second equation in the system of moment equations are coupled and necessary. However, in order to facilitate comparison with previous work, this study still adopts a three-equation model for the particle fields. Initially, the lower stream is free of particles, while the upper stream is populated by nanoparticles. The initial $k$th moment for $k = 0,1,2$ are $M_{00} = 1, M_{10} = 1, M_{20} = 4/3$. The flow field and particle field can be solved using the numerical method listed in **Appendix II**.

**Results and Discussion**

**The self-similarity of unperturbed spatial mixing layer**

The case of the two-dimensional spatial mixing layer was considered in the domain of $0 \le x \le 24\pi$ and $-2\pi \le y \le 2\pi$ (or $L_x = 24\pi\ and\ L_y = 4\pi$). The velocity ratio for this case was set at $U_1/U_2 = 2$. This value represents a moderately sheared layer intermediate between the case of a single-stream mixing layer and the splitter-plate wake case. The Reynolds number is set at $Re = 200$. The domain was discretized using 1536 points to represent the streamwise extend of the domain; 256 collocation points were used to represent the transverse direction. A time step of 0.01 was used in this work. In the absence of external forcing, the distributions of vorticity are shown in **Figure 2a**. According to the definition of vorticity,

$$\Omega = \frac{\partial U}{\partial y} - \frac{\partial V}{\partial x} \tag{50}$$

It can be seen that the thickness of the mixing layer continues to increase along the streamwise direction. **Figure 3** reveals the relationship between the momentum or vorticity thickness and the streamwise location for the unperturbed mixing layer. The momentum thickness ($\theta$) and vorticity thickness ($\delta_\Omega$) are defined as

$$\theta = \int_{-\frac{L_y}{2}}^{\frac{L_y}{2}} \frac{(U_1 - U)(U - U_2)}{(U_1 - U_2)^2} dy \tag{51}$$

$$\delta_\Omega = \frac{U_1 - U_2}{\left(\frac{dU}{dy}\right)_{max}}; \tag{52}$$

A square root relationship fit to these computed results are $\theta = 0.0467\sqrt{x + 0.5971}$ and $\delta_\Omega = 0.2041\sqrt{x + 0.5440}$, respectively. The spatial mixing layer is responded with the classical laminar, square root growth characteristics, which is evident by the close agreement indicated by the computed results. The mean field statistics for the streamwise and transverse velocity component and vorticity are illustrated in **Figure 4**. Clearly, these results are representative of a self-similar layer. Although the distribution of transverse velocity is not completely coincident, it also exhibits similarity with the amplitude continuous decreasing with the streamwise location.

The resolution of a fluid flow simulation, defined by the fineness of computational grid or the number of discrete points used to represent the flow domain, has a significant impact on the accuracy and reliability of the results. When simulating fluid flows, particularly shear flows in a mixing layer, the grid resolution plays a crucial role in accurately capturing the dynamics of vortices and other flow structures. Usually, DNS requires resolving all scales of motion down to the Kolmogorov scale, and the number of grid points ($N$) needed in each dimension scales with the Reynolds number as $N \sim Re^{3/4}$. The computational times for $768 \times 128$, $1536 \times 256$ and $3076 \times 512$ grids are 2013.752, 12472.312 and 47279.074 seconds, respectively. the results indicating that the computational cost increases exponentially as the number of grids increases. **Figure 3 also** shows the vorticity thickness for different resolution at a Reynolds number of 200. For moderate Reynolds numbers, the results obtained from the three selected grids in this article have good consistency. In order to balance computational accuracy and efficiency, the number of grids used in this article is $1536 \times 256$.

**Eigenvalue and eigenfunction based on linear stability theory**

It is not difficult to see that the pseudo-spectral method describes the evolution characteristics of small disturbance in both temporal and spatial patterns, with the same phase velocity expression, but they also have significance differences. For the temporal mode, it is actually very similar to the dynamical system of ordinary differential problems, that is, it describes the evolution of the spatial mode of disturbance over time. From this point of view, although the linear stability theory localizes the expression of disturbances, their instability characteristics still have integrity, that is, disturbances at different flow positions have the same dynamic characteristics. On the contrary, the instability characteristics described by spatial patterns are actually localized. At a given disturbance frequency, the disturbance propagates from upstream to downstream, and its growth rate actually corresponds to the amplitude amplification rate of each local station. Seeing this, it is not difficult to see the importance of the parallelism assumption mentioned in the **Appendix I**.

Only when the basic flow is parallel (or invariant) along the direction of flow, can the growth rate be accurately used to define the growth rate of disturbance waves. In other words, only under the premise of parallel basic flow, can the eigenvalues and eigenfunctions given by the linear stability theory accurately represent the true meaning of eigenvalues and eigenfunctions. However, the vast majority of flows, such as spatial mixing layer flows, are non-parallel along the flow direction, and only change slowly along the flow direction. Therefore, the assumption of parallelism actually ensures that we can localize linear stability problems and obtain an approximate solution under local parallelism condition, which ignores the gradient of the fundamental flow along the direction of flow.

From the derivation of the linear stability equation as shown **Appendix I**, using the assumption of parallelism to determine the local eigenvalues and eigenfunctions also means that the gradient of the perturbed eigenfunctions along the flow direction is ignored. Therefore, for the stability problem of a general shear flow that develops along space, we can solve an eigenvalue problem that introduces the assumption of parallelism at each flow direction station to obtain the disturbance characteristics of the entire field. in other words, by introducing the assumption of parallelism, a partial differential equation is locally differentiable. It is precisely the variation of different physical quantitates along the flow direction may not be completely consistent because of the non-parallel features.

Obviously, if our goal is to study the transition process of shear flows, spatial models are more suitable. Because the disturbances in these flow processes happen to gradually develop from upstream to downstream and trigger transition. What can be measured in the experiment is also the evolution of disturbances in spatial mode. It precisely for this reason that spatial patterns have a special place in the study of transition problems. Of course, this does not mean that time patterns are not important. In fact, when studying some basic physics problems, if the fundamental flow is parallel, it is obviously simpler and more memory efficient to use time mode for research. After all, for parallel flows, the two patterns can be transformed into each other through the Gaster transform. **Figure 5** show the distribution of eigenvalue for spatial models and the corresponding eigenfunctions under eigenvalue with maximum imaginary part. The eigenfunctions will serve as the inlet boundary and initial conditions of the forced spatial mixing layer simulation.

**Analysis of forced mixing layer flow**

The case of forcing mixing layer considered in this investigation applies a set of time dependent perturbations to the velocity component at the inlet plane. These perturbations are generated by the means of linear stability analysis for the most unstable mode. **Figure 2b** provides a snapshot of the vorticity field at the simulation. Sufficient time has been elapsed for the initial, start-up field to have wash-out of the computational domain (the dimensionless time is $t = 300$). From the snapshots of vorticity in the forced mixing layer, it can be seen that there is a clear spatial evolution of vortex entrainment, rolling, paring, merging and dissipation processes compared to the unperturbed mixing layer. **Figure 6** compares the computed streamwise growth of thickness in forced mixing layer with the laminar growth of thickness in the unperturbed mixing layer. Note that, the inlet perturbation has strong influence on the growth of mixing layer. In the growth stage of vortex generation near to the inlet flow, the influence of perturbation on the thickness is small, while in the stage of full development of vortex, the quasi-periodic oscillation of the thickness along the streamwise direction becomes more pronounced. The entrainment, rolling and pairing, emerging of vortices significantly increase the thickness of the mixing layer. Compared to laminar flow, the resolution also has affects the streamwise growth of thickness in forced layer. **Figure 7** reveals the effect of resolution on the streamwise growth of the momentum and vorticity thickness in the forced mixing layer at Reynolds number 200. The resolution mainly affects the phase change of oscillations. The reason for this difference is that the eigenvalue and its corresponding eigenfunction are related to the Reynolds number and grids number. If statistical averaging is performed, this phase variation can be ignored, just like the evolution of the thickness in the laminar flow. this speculation will be supported by **Figure 8**. Figure 8 illustrate time traces of the results for the velocity components at the selected location in the forced mixing layer. The results clearly indicate that the response of the layer is very periodic. This is due to the periodic forcing imposed at the inlet flow of the layer. The peak-to-peak time lapse in these curves provides evidence of the passage of a structure. The time laps together with an assumed mean advection speed for these structures allows estimation of the scale of a structure. The amplitude of disturbance is highest in the middle of the flow field, while it is lower in the initial and developed stages of the flow field. Therefore, the spatial evolution of the flow field can be divided into three parts, i.e., initial region ($0 \leq x \leq 4\pi$), transition region ($4\pi \leq x \leq 16\pi$) and developed region ($x \geq 16\pi$). In the initial region, the perturbations are small, and the characteristics of the flow field roughly the same as the laminar flow. In the transition region, perturbations begin to grow, the strongest and most organized vortical structures forms, which appear as circular/elliptical patterns. The intensity of perturbations decreases away from the centerline. In the fully developed region, perturbations gradually decays, size of the vortices grows slightly downstream, vortex structure maintain similar shape and spacing, and the vortical structure becomes blurred. The results can also be found through the spatial distribution of turbulence intensity as shown in **Figure 9**. This distribution is typical of well-developed spatial mixing layer showing Kelvin-Helmholtz instability and subsequent vortex formation. Therefore, the selected domain simulated in this article accurately reflects the characteristics of the spatial mixing layer, which lays the foundation for the numerical simulation of the evolution of particle field.

**The spatial distribution of particle moments**

The distribution pattern is characteristic of particle dispersion in a spatial mixing layer as shown in **Figure 10**, which reveals how the initially concentrated particles are gradually mixed and dispersed by the flow structures. Compared with the distribution of vorticity in the flow field, the distribution of the three moments in the particle field is similar to the distribution of vorticity as shown in **Figure 2b**, which demonstrate the influence of large-scale structures on particle distribution. The spatial distribution of particle moments can be divided into two parts, i.e., one is the mean particle moments, and the other is perturbated part, which can be expressed as:

$$\begin{aligned} M_0 &= \overline{M_0} + M_0' \\ M_1 &= \overline{M_1} + M_1' \\ M_2 &= \overline{M_2} + M_2' \end{aligned} \tag{53}$$

Where the mean particle moments are defined as

$$\begin{aligned} \overline{M_0} &= \frac{1}{T}\int_0^T M_0(x,y,t)dt \\ \overline{M_1} &= \frac{1}{T}\int_0^T M_1(x,y,t)dt \\ \overline{M_2} &= \frac{1}{T}\int_0^T M_2(x,y,t)dt \end{aligned} \tag{54}$$

**Figure 11** shows the spatial distribution of mean particle moments at the conditions as $Re = 200, Sc = 1, Da = 1$. The distributions of the mean particle moments are consistent with the vorticity distribution in the unperturbed mixing layer as shown in **Figure 2a**. **Figure 12** shows the spatial distribution of square-root of fluctuating particle moments, which exhibits a wave-like pattern in the mixing region; the distributions of fluctuating particle moments are more similar to the structure of vortex in the forced mixing layer. The similar distributions between vorticity and particle moments in the forced mixing layer can be explained by several key reasons:

1. Fluid-particle interaction: vortical structures directly influence particle motion, while particle tend to follow fluid elements due to the small Stokes number of nanoparticles;
2. Entrainments mechanism: the vortices act as primary mixing mechanisms, and both vorticity and particles are entrained by the same flow structures, and have the similar patterns of spreading and organization. Kelvin-Helmholtz instability drives the vortex formation and controls particle dispersion pattern; and shear layer dynamics influence the distribution of both vorticity and particle moments.
3. Coherent structures: The spiral-like patterns of fluctuating distribution of particle moments indicates the vortex formation;
4. Physical interpretation: the similarity shows how particle concentration fluctuates around the mean, demonstrates the strong coupling between flow structures and particle distribution, reveals the spatial organization of nano-particle laden mixing flow, particularly in the transition region where initial coherent structure occurs.

**The evolution of particle moment along the streamwise direction**

Coagulation is a process where particles collide and stick together, forming larger particles, which affects the particle size distribution and, consequently, the moments of the particle size distribution. In a spatial mixing layer, the evolution of particle moments due to coagulation can be significantly influenced by the flow dynamics. Since the mean velocity follows the streamwise direction, which means that the evolution of particle moments along the streamwise direction is the evolution of particle moments over time. **Figure 13** show the distribution of particle moments at different lines along the transverse direction in the mixing layer under the conditions $Re = 200, Sc = 1, Da = 1$ and $t = 300$. Overall, according to the distribution of particles in space, the particle fields can be roughly divided into three regions.

In the upper half of the mixing layer ($L_y/8 \leq y \leq L_y/2$), the flow field has not been affected by the large-scale coherent structures, and the evolution of particles is basically consistent with the 0-dimensional problems, i.e., the zeroth order particle moment decreases exponentially over time, the first order particle moment almost remains constant, and the second particle moment increases exponentially over time.

In the central line of the mixing layer, the large-scale structure has not been formed at the initial stage, the particle number density undergoes a growth process under the combined action of coagulation, diffusion and convection. During the rolling, pairing and merging stages of vortices, the evolution of particle moments exhibits significant fluctuations due to the effect of flow dynamics. In the dissipation stage of the vortex, the effect of the flow dynamics gradually disappears, and the evolution of particles returns to the 0-dimensional problem.

In the lower half of the mixing layer ($-L_y/2 \leq y \leq -L_y/8$), there are no particles at the bottom of the flow field in the initial stage. Under the action of convection and diffusion, particles are gradually transported from the upper part to the lower part of the flow field. Therefore, the particle number density shows a continuous growth process. At the same time, due to the effect of large-scale eddies, some fluctuations are also present.

**The evolution of particle moment along the transverse direction**

Due to the mean velocity in the transverse direction is zero, the evolution of the mean particle moments in the transverse direction mainly focuses on the effect of diffusion. At the same time, the large-scale structure has significantly influenced by the evolution of the particle moment. The evolution of the particle moments in the transverse direction can be decomposed into mean field and pulsating field. the mean particle field reflects the diffusion effect, while the pulsating particle field reflects the influence of large-scale structure on the particle field. **Figure 14** presents nine subplots illustrating the evolution of particle moments along the transverse direction in a spatial mixing layer under specific flow conditions: Reynolds number $Re = 200$, Schmidt number $Sc = 1$, and Damköhler number $Da = 1$. These plots provide a comprehensive analysis of particle distribution characteristics through total, mean, and fluctuating components of various particle moments ($M_0, M_1, M_2$). Overall, the particle number density shows a gradually decreasing along the streamwise direction, while the particle volume concentration remains constant, and the dispersity shows an increasing trend.

The first row (subplots a-c) focuses on particle number density distributions: Subplot (a) shows the total particle number density ($M_0$), displaying a characteristic S-shaped profile that transitions from high concentration in the upper stream to low concentration in the lower stream. The curves for different streamwise locations demonstrate the progressive mixing process, with the transition region broadening downstream while maintaining consistent asymptotic values. Subplot (b) presents the mean particle number density ($\overline{M_0}$), revealing the time-averaged concentration distribution that shows a smoother transition between the streams and highlights the stable features of the mixing process. Subplot (c) illustrates the fluctuating particle number density (${M_0}'$), which peaks in the mixing region where concentration gradients are strongest, indicating the regions of most intense particle-vortex interaction.

The second row (subplots d-f) examines particle volume concentration: Subplot (d) displays the total particle volume concentration ($M_1$), showing similar S-shaped characteristics to the number density but with distinct features in the transition region that reflect the combined effects of particle number and size variations. Subplot (e) presents the mean particle volume concentration ($\overline{M_1}$), demonstrating how the average volume distribution evolves across the mixing layer and providing insights into the steady-state particle volume distribution. Subplot (f) shows the fluctuating particle volume concentration (${M_1}'$), revealing regions where particle volume concentrations deviate most significantly from their mean values and highlighting areas of enhanced mixing and particle clustering.

The third row (subplots g-i) addresses the dispersity of particle size distribution: Subplot (g) illustrates the dispersity of particle size distribution ($M_2$), showing how the spread of particle sizes varies across the mixing layer and evolves downstream. This distribution provides crucial information about the uniformity or heterogeneity of particle sizes in different regions of the flow. Subplot (h) presents the mean dispersity of particle size distribution ($\overline{M_2}$), revealing the time-averaged characteristics of particle size variations and demonstrating the stable features of size distribution across the mixing layer. Subplot (i) shows the fluctuating dispersity of particle size distribution (${M_2}'$), indicating regions where particle size distributions experience the most significant temporal variations and highlighting the dynamic nature of particle size evolution.

Throughout all subplots, several consistent features are observable:

- The profiles for total and mean quantities maintain smooth S-shaped transitions between the upper and lower streams;
- Fluctuating components show peak values in the mixing region where gradients are strongest;
- The transition regions generally broaden with downstream distance, indicating progressive mixing;
- The upper and lower stream values remain relatively constant, preserving the boundary conditions
- Different moments show varying sensitivity to the mixing process, revealing the complex nature of particle-fluid interactions

The evolution patterns demonstrate that particle transport and mixing in the spatial mixing layer is a multifaceted process, with different moments exhibiting distinct behaviors. The presentation of total, mean, and fluctuating components for each moment provides a complete picture of both steady-state and dynamic aspects of particle behavior. The total quantities show the instantaneous state of the system, the mean quantities reveal the time-averaged behavior and stable features, while the fluctuating components highlight regions of dynamic activity and intense mixing. The consistent format across all subplots, with multiple streamwise locations plotted together, enables direct comparison of how different particle characteristics evolve and interact throughout the mixing layer. This comprehensive view reveals the complex interplay between particle transport, mixing processes, and flow structures. The analysis of number density, volume concentration, and size dispersity through their total, mean, and fluctuating components provides deep insights into the fundamental physics governing particle-laden mixing layers. The results suggest that particle dynamics in the mixing layer are characterized by strong spatial variations and temporal fluctuations, particularly in the mixing region where the interaction between particles and flow structures is most intense. The evolution of these distributions downstream indicates a progressive organization of the particle-laden flow, with initial sharp gradients giving way to more gradual transitions as mixing proceeds. This detailed characterization of particle moment evolution is essential for understanding and predicting the behavior of particle-laden flows in practical applications and for developing improved models for particle transport and mixing processes.

**The evolution of particle size distribution along the streamwise direction**

Advection is the transport of particles and vortices by the bulk motion of the fluid, this means that particles move with the flow, maintaining their relative positions and structures as they are carried by the fluid. This is the main reason for the similarity of the distribution between the particle moments and the vortices. Advection, interacting with shear layers and coherent structures, contribute to processes such as vortex stretching, tilting, merging and splitting. These interactions play a crucial role in the energy cascade and overall dynamics of flows as shown in **Figure 10**. While advection primarily affects the spatial distribution of particles, it can indirectly influence the particle size distribution under certain conditions. Advection moves particles along with the fluid flow, redistributing them spatially within the flow domain. This process does not direct change the size of individual particles but affects where particles of different size are located. By redistributing particles, advection can bring particles of different sizes into closer proximity, potentially increasing the collision rate. This can enhance coagulation leading to changes in the particle size distribution.

Diffusion can create or modify concentration gradients of particles within the flow, which can influence local particle interactions and processes such as coagulation. When combined with coagulation, diffusion can significantly impact of the evolution of the particle size distribution. The diffusion term in the particle PBE can influence the spatial distribution of particles, which in turn affects the local coagulation rates. Advection can work in conjunction with diffusion to spread particles throughout the flow domain. The combined effect of advection and diffusion can lead to a more uniform distribution of particles, affecting the overall size distribution. Overall, in the mixing layer, advection and diffusion can move particles from regions of high concentration to regions of low concentration. This redistribution can lead to increased local particle concentrations and higher collision rates, resulting in changes to the particle size distribution over time.

**Figure 15** illustrated the distribution of particle size dispersity ($M_C$) across the vertical direction of a mixing layer at different streamwise locations. The profiles show distinct characteristics as they evolve downstream. In the upper stream ($y > \pi/2$), all curves converge to a constant value of approximately 2.0, indicating a uniform particle size distribution. In the central region ($-\pi/2 \leq y \leq \pi/2$), the evolution of dispersity is mainly dominated by the large-scale vortex structures. In the lower stream, the dispersity starts from zero, suggesting initially no particles. The most interesting features appear in the mixing region ( $-2\pi \leq y \leq -\pi/2$ ), where each streamwise location exhibits a peak in dispersity, with maximum values reaching around 3.5-4.0. These peaks shift progressively upward and become more pronounced as the flow develops downstream, indicating enhanced particle size variation in the mixing region. The location and magnitude of these peaks suggest that the mixing process creates regions of high particle size heterogeneity, likely due to the interaction between particles and flow structures. The gradual transition between the upper and lower streams becomes broader downstream, reflecting the growing influence of mixing processes on particle size distribution. This evolution pattern demonstrates how the mixing layer significantly affects the local diversity of particle sizes, particularly in the shear region where flow interactions are strongest.

In the fully developed stage of the mixing layer, the intensity of vortices gradually decreases. The results in self-similar coordinate as shown in **Figure 4** are also indicate that the time averaged statistics for velocity, vorticity and turbulence intensities and Reynolds stress distribution tend to collapse at the flow downstream locations as shown **Figure 9.** If the simulation range is large enough, it will be seen that the mixing layer flow will develop into Couette flow. Under these conditions, the particle distribution in the horizontal direction is uniform, or the gradient of particle moments is zero, i.e., $\partial n/\partial x = 0$. In the vertical direction, the distribution of particle moments is basically linear, which means that the second derivative of particle moments in the vertical direction is zero. Then the diffusion term in the PBE can be eliminated. Together with the asymptotic distribution of velocities in **Figure 8** ($V{\sim}0$), the advection term can also be eliminated. And the particle population balance equation degenerates into the Smoluchowski coagulation equation, i.e.,

$$\frac{\partial n}{\partial t} = \left[\frac{\partial n}{\partial t}\right]_{coag} \tag{55}$$

Therefore, the particle size distribution under asymptotic conditions is consistent with the 0-dimenaional problem as shown in **Figure 16**.

**The effect of Schmidt number**

The Schmidt number ($Sc$) is a dimensionless quantity that characterizes the ratio of momentum diffusivity (viscosity) to mass diffusivity (diffusion) in a fluid. It plays a significant role in determining the behavior of particles in a fluid flow, particularly in processes involving diffusion and mixing. When $Sc$ is much less 1, the mass diffusivity is much greater than the kinematic viscosity. This means that particles diffuse rapidly compared to the rate at which momentum diffuses. In this regime, the rapid diffusion of particles leads to a more uniform distribution of particles in the fluid. Coagulation and other particle interactions are influenced by the enhanced mixing and dispersion. When $Sc$ is much greater than 1, the mass diffusivity is much less than the kinematic viscosity. This means that particles diffuse slowly compared to the rate at which momentum diffuses. In this regime, the slow diffusion of particles leads to higher local concentrations of particles, increasing the likelihood of collisions and coagulation. When $Sc$ is around 1, the mass diffusivity and kinematic viscosity are comparable. This means that the rate of particle diffusion and momentum diffusion are balanced. In this regime, both diffusion and coagulation processes significantly influence the evolution of particle moments. The particle size distribution evolves due to a combination of mixing and particle interactions. The moments evolve in a more complex manner as shown in **Figure 17**, with both lower order and higher order moments changes due to the interplay between diffusion and coagulation. The influence of Schmidt number revolves around a baseline vibration, the larger the Schmidt number, the greater the fluctuation amplitude. In addition, the reciprocal of the product of Schmidt number and Reynolds number in the diffusion term of particle evolution equation. therefore, the effect of Schmidt number on the evolution of particles is equivalent to the effect of Reynolds number.

For convenience, when discussing the influence of Schmidt number (Sc), the main focus is on analyzing the distribution patterns of particle moments in the horizontal and vertical directions at the center of the flow field. **Figure 17** in the left shows the distribution of particle field in the central horizontal direction; **Figure17** in the right shows the distribution of the particle field in the central vertical direction. In the horizontal direction (left column), the top subplot reveals a decaying oscillatory pattern of particle number density, where different Schmidt numbers affect the decay rate and fluctuation intensity. Higher Schmidt numbers tend to maintain stronger fluctuations over longer distances, indicating reduced diffusive damping. The middle subplot exhibits regular oscillations of particle volume concentration with varying amplitudes, showing how Schmidt number influences the particle-vortex interactions and the resulting periodic fluctuations in particle distribution. The bottom subplot demonstrates growing oscillations of the dispersity of particle size distribution, where Schmidt number affects both the growth rate and wavelength of the fluctuations.

The vertical distributions (right column) show characteristic S-shaped profiles typical of mixing layers. The top right subplot displays the particle number density distribution across the layer, where Schmidt number notably affects the thickness of the transition region. Higher Schmidt numbers typically result in sharper gradients and thinner mixing regions due to reduced diffusive mixing. The middle and bottom right subplots present similar S-shaped transitions for different particle moments, with Schmidt number clearly influencing the spreading rate and mixing layer thickness. Due to the interaction between coagulation and diffusion, the maximum value of dispersity near the interface (**Wang et al., 2012**). Key features across all plots include: Higher Schmidt numbers generally lead to sharper gradients in vertical profiles, more persistent fluctuations in the streamwise direction, thinner mixing layers, and stronger particle-flow coupling. Lower Schmidt numbers result in more diffuse transitions, faster decay of fluctuations, broader mixing regions and enhanced mixing effects. The combined analysis of horizontal and vertical distributions provides a comprehensive view of how Schmidt number affects both the dynamic evolution and spatial distribution of particle moments in the mixing layer. This demonstrates the fundamental role of the Schmidt number in determining the balance between convective and diffusive transport processes in particle-laden mixing layers.

**The effect of Damkohler number**

The Damkohler number ($Da$) is a dimensionless quantity that characterizes the relative importance of particle dynamics to transport processes (such as advection, diffusion, etc.) in a system. In the context of particle laden flows, the Damkohler number can significantly influence the evolution of particle moments, especially when coagulation processes are involved. When Damkohler number is much less than 1 ($Da \ll 1$), the transport processes dominate over the coagulation. The evolution of particle moments is primarily governed by the mixing and diffusion of particles. Coagulation occurs slowly, and the particle size distribution changes gradually over time. When Damkohler number is much greater than 1 ($Da \gg 1$), the coagulation process dominates over the transport processes. In this regime, coagulation occurs rapidly, leading to significant changes in the particle size distribution. Particles collide and coalesce quickly, forming larger particles. When Damkohler number is around 1 ($Da \sim 1$), the coagulation and transport processes are balanced. In this regime both coagulation and mixing influence the evolution of particle moments. The particle size distribution evolves due to a combination of coagulation and transport effects as shown in **Figure 18**. The Damkohler number has almost no effect on the distribution of the particle volume concentration, which can also be seen from the source term of the first order moment equation. due to the volume conservation of coagulation, the source term is 0. In addition, the spatial distribution of particle moments is similar to that of vortices, and Damkohler number mainly affects the mean and amplitude of zeroth and second order moments.

**Figure 18** illustrates the profound influence of Damköhler number ($Da$) on the evolution of particle moments through six interconnected subplots, divided between streamwise evolution (left column) and vertical distributions (right column). In the streamwise direction (left column), the evolution patterns reveal complex interactions between coagulation and transport processes. The top subplot demonstrates a decaying oscillatory behavior of particle number density ($M_0$), where higher $Da$ values accelerate the decay of fluctuations, indicating stronger coagulation effects in dampening particle moment variations. The middle subplot exhibits fascinating periodic oscillations of particle volume concentration whose amplitudes and frequencies are significantly modulated by $Da$, showcasing the intricate balance between coagulation kinetics and flow dynamics. The bottom subplot reveals how coagulation rates affect the growth of the dispersity of particle size distribution ($M_2$), with higher $Da$ values leading to distinct patterns in the development of particle second moment fluctuations.

The vertical distributions (right column) present equally compelling evidence of $Da$ influence on spatial organization. The top subplot shows characteristic S-shaped profiles whose shapes and transition regions are markedly affected by coagulation rates, with higher $Da$ values typically resulting in modified gradient patterns. The middle and bottom subplots further demonstrate how coagulation reshape the vertical structure of particle moments, with different $Da$ values leading to distinct stratification patterns and mixing characteristics.

The combined analysis reveals several critical features of coagulating mixing layers. Higher Damköhler numbers generally lead to more rapid attenuation of particle number density, reflecting the dominant role of coagulation in shaping particle distributions. Lower $Da$ values, conversely, allow transport processes to play a more significant role, resulting in different evolution patterns and spatial organizations. This interplay between coagulation and transport processes, quantified by the Damköhler number, fundamentally determines the nature of particle-laden coagulating mixing layers. These observations have significant implications for understanding and predicting the behavior of coagulating particle-laden flows. The strong dependence of particle moment evolution on $Da$ suggests that careful consideration of coagulation rates relative to transport timescales is crucial for accurate modeling and control of such flows. The complex patterns observed in both spatial evolutions highlight the need for sophisticated approaches in analyzing and predicting the behavior of coagulating mixing layers.

**Conclusions**

The two dimensional incompressible, spatially developing, forcing mixing layer has been simulated in this work. The inflow boundaries are excited to generate the forced mixing layer simulations. And advection type outflow boundary condition was employed in this work. This condition appears to allow a simulation that does not distort the structures as they exit the computational domain. This simulation reflects the imposition of a time dependent perturbation function at the inlet flow. this perturbation corresponding to the fundamental mode corresponding to solutions to the Orr-Sommerfeld equation for the hyperbolic tangent inviscid shear layer profile. The results of the simulation capture the physics of the forced mixing layer quite well.

When nanoparticles are present in this flow, their evolution is influenced by various physical processes and the unique characteristics of the mixing layer. At the start, nanoparticles are typically distributed non-uniformly across the mixing layer, concentrated in the upper of the fluid streams. As the mixing layer develops, large-scale vortices form due to Kelvin-Helmholtz instabilities, which is captured through direct numerical simulation and linear stability analysis. These vortices enhance mixing, causing nanoparticles to be transported across the layer. Advection by the mean flow and large eddies redistributes particles throughout the mixing region, this enhances mixing and leads to a more uniform distribution over time. Brownian motion causes nanoparticle to diffuse, gradually spreading them across the mixing layer. The diffusion rate depends on particle size, with smaller particles diffusing more rapidly. To simplify the calculation, a moment based average diffusion coefficient model is proposed. Increased particle interactions can lead to coagulation, larger particles may form as smaller nanoparticles collide and stick together, the AK-iDNS framework is used to approximate the Smoluchowski coagulation equation. Strong velocity gradients in the mixing layer can cause shear-induced migration of particles, which can lead to non-uniform spatial distributions and preferential concentration in certain regions of the flow. Over time, a more homogeneous distribution is approached in the horizontal direction. The particles exhibit similar asymptotic behavior at that of 0-dimensional problem. However, due to the non-uniform spatial distribution of particles in the vertical direction, there are some differences in the distribution of the dimensionless particle moment. But the evolution of particle moments has almost the same scaling growth rate.

The investigation of nanoparticle evolution in spatial mixing layers reveals several key findings regarding the interaction between mixing layer flow and Brownian coagulation. The results demonstrate that the spatial heterogeneity of particle concentrations, coupled with vortex transport mechanisms, creates distinct regions of enhanced coagulation activity. The shear layer's role in particle transport and mixing proves crucial in determining the final particle size distribution and spatial arrangement. Local vorticity characteristics significantly influence coagulation rates, with higher turbulence intensities generally leading to accelerated particle growth. These findings have important implications for industrial applications involving nanoparticle synthesis and atmospheric aerosol dynamics. Future work should focus on extending this analysis to more complex flow configurations and incorporating additional physicochemical processes that may affect particle evolution. The insights gained from this study contribute to our understanding of particle-vorticity interactions and provide a foundation for improving models of aerosol dynamics in engineering applications.

**Acknowledgements**

This work was funded by the National Natural Science Foundation of China with grant number 11972169.

**Nomenclature**

| | |
|---|---|
| $\rho$ | fluid density |
| $p$ | pressure |
| $p'$ | disturbances of pressure |
| $\nu$ | kinematic viscosity of fluid |
| $u$ | velocity component in the $x$ direction |
| $v$ | velocity component in the $y$ direction |
| $U$ | base flow in in the $x$ direction or the characteristic velocity |
| $V$ | base flow in in the $y$ direction |
| $u'$ | disturbances in in the $x$ direction |
| $v'$ | disturbances in in the $y$ direction |
| $U_1$ | far field velocity at the upper flow |
| $U_2$ | far field velocity at the lower flow |
| $x$ | the Cartesian coordinate in the stream-wise direction |
| $\boldsymbol{y}$ | the Cartesian coordinate in the transverse direction |
| $\theta$ | the initial momentum thickness |
| $n$ | the number density of particles |
| $v$ | particle volume |
| $v_a$ | the algebraic mean volume |
| $t$ | time |

$\beta$ the collision kernel of coagulation
$\bar{\beta}$ the average collision kernel
$k_B$ the Boltzmann's constant
$T$ the fluid temperature
$\rho_p$ the particle density
$D_n$ the particle diffusion coefficient
$f$ the friction coefficient
$m$ the molecular mass of fluid molecules.
$\alpha_p$ the accommodation coefficient,
$d_p$ the diameter of particles
$Da$ the Damkohler number

$Re$ the Reynolds number
$St$ the Stokes number
$Sc_M$ the Schmidt number based on the particle moment
$L$ the characteristic length
$\alpha$ the scaler factor or coefficient
$\alpha_p$ the accommodation coefficient
$p$ the proportional factor
$q$ the power index
$M_0$ total particle number concentration
$M_1$ the particle volume concentration
$M_2$ the dispersity of particle size distribution
$M_k$ $k$th order moment of the volume-based particle size distribution
$M_{k0}$ the initial value of $k$th order moment
$\eta$ the dimensionless particle volume,
$\psi$ the dimensionless particle size distribution
$\mathcal{L}$ the Laplace operator,
$s$ the Laplace transform parameter,
$\Psi$ the transformed dimensionless particle size distribution
$A$ the total collision frequency
$C$ the gain term
$g$ the loss term
$\varepsilon$ error limits
$k$ the spatial wavenumber in $x$ direction
$\omega$ the temporal pulsation
$a, b, c, d$ the coefficient
$\phi$ the stream function
AK average kernel method
DNS direct numerical simulation
iDNS iterative direct numerical simulation
PSD particle size distribution

**Appendix I: computation of spatial stability for parallel flows**

The Kelvin-Helmholtz instability is the primary mechanism for initial growth, which leads to the formation of large-scale vortical structures. The theory of instability of parallel laminar flows decomposes the motion into a mean flow and a disturbance superimposed on it (**Schmid and Henningson, 2001; Drain, 2002**). Let the mean flow, which may be regarded as steady, be described by its Cartesian velocity component $U$, $V$ and pressure $P$. The corresponding quantities for the non-steady disturbance will be denoted by $u'$, $v'$ and $p'$, respectively.

$$\begin{cases} u = U + u' \\ v = V + v' \\ p = P + p' \end{cases} \tag{AI.1}$$

The Navier-Stokes equations linearized about the base flow profile $U$ in two-dimensions is

$$\begin{cases} \frac{\partial u\prime}{\partial x} + \frac{\partial v\prime}{\partial y} = 0\text{。} \\ \frac{\partial u\prime}{\partial t} + U\frac{\partial u\prime}{\partial x} + v'\frac{dU}{dy} = -\frac{\partial p'}{\partial x} + \frac{1}{Re}\Delta u' \\ \frac{\partial u\prime}{\partial t} + U\frac{\partial v\prime}{\partial x} = -\frac{\partial p\prime}{\partial y} + \frac{1}{Re}\Delta v' \end{cases} \tag{AI.2}$$

where $\Delta$ is the Laplacian operator. Rewriting this equation in operator form as

$$\left[-\begin{pmatrix} \partial_t & 0 & 0 \\ 0 & \partial_t & 0 \\ 0 & 0 & 0 \end{pmatrix} + \begin{pmatrix} -U\partial_x + \Delta/Re & dU/dy & -\partial_x \\ 0 & -U\partial_x + \Delta/Re & -\partial_y \\ \partial_x & \partial_y & 0 \end{pmatrix}\right]\begin{pmatrix} u' \\ v' \\ p' \end{pmatrix} = 0 \tag{AI.3}$$

where $\partial_\{x, y, t\}$ denote the partial derivative operators. We now consider normal modes, that is we assume an oscillating behavior in $x$ and time of the flow solution

$$\begin{cases} u' = u'(y)\exp[i(kx - \omega t)] \\ v' = v'(y)\exp[i(kx - \omega t)] \\ p' = p'(y)\exp[i(kx - \omega t)] \end{cases} \tag{AI.4}$$

where $k$ is the spatial wavenumber in $x$ direction and $\omega$ is the temporal pulsation. The exponential structure allows the solution to oscillate and grow/decay in space and time, depending on the real and imaginary parts of $k$ and $\omega$. In the temporal analysis, the solution grows/decay and oscillates in time but only oscillate in space: $k \in R$ is given, and one obtains $\omega$ from the dynamic equations. In the spatial analysis, one assumes that the solution only oscillates in time at a given spatial position, but is allowed to grow/decay and oscillated in space: $\omega \in R$ is given, and $k$ is obtained from the dynamic equations.

Injecting the normal modes in the dynamic equations, we can replace

$$\begin{cases} \partial_t = -i\omega \\ \partial_x = ik \\ \partial_{xx} = -k^2 \end{cases} \tag{AI.5}$$

And it is obtained

$$\omega EX = (A_0 + A_1 k + A_2 k^2)X \tag{AI.6}$$

Where $E$, $A_0$, $A_1$, $A_2$, $X$ are represented as follows

$$E = \begin{pmatrix} \partial_t & 0 & 0 \\ 0 & \partial_t & 0 \\ 0 & 0 & 0 \end{pmatrix};$$

$$A_0 = \begin{pmatrix} \partial_{yy}/Re & -dU/dy & 0 \\ 0 & \partial_{yy}/Re & -d/dy \\ 0 & d/dy & 0 \end{pmatrix};$$

$$A_1 = \begin{pmatrix} -iU & 0 & -i \\ 0 & -iU & 0 \\ i & 0 & 0 \end{pmatrix};$$

$$A_2 = \begin{pmatrix} -I/Re & 0 & 0 \\ 0 & -I/Re & 0 \\ 0 & 0 & 0 \end{pmatrix};$$

$$X = \begin{pmatrix} u \\ v \\ p \end{pmatrix}$$

This equation is the dispersion relation. If $k$ is given, $\omega$ is the eigenvalue of a generalized eigenvalue problem; whereas if $\omega$ is given, $k$ is the eigenvalue of a polynomial eigenvalue problem. In general, the direct solution of polynomial eigen value problems can be heavy. For the present case, it can transform the polynomial eigenvalue problem into a generalized eigenvalue problem. To do this, we argument the system with the variable

$$Y = kX \qquad \text{(AI.6)}$$

The dispersion equation can be simplified as

$$\begin{pmatrix} -\omega E + A_0 & 0 \\ 0 & I \end{pmatrix}\begin{pmatrix} X \\ Y \end{pmatrix} + k\begin{pmatrix} A_1 & A_2 \\ -I & 0 \end{pmatrix}\begin{pmatrix} X \\ Y \end{pmatrix} = 0 \qquad \text{(AI.7)}$$

By way of example, the boundary conditions for the spatial mixing layer demand that the components of the perturbation velocity must vanish at a large distance from the interface (free stream).

$$u' = 0, and\ v' = 0\ at\ y = \pm\infty \qquad \text{(AI.8)}$$

## Appendix II: numerical method for fluid flow

The general approach of the simulation of fluid flow is described as follows. While $u, v, p$ are the solutions to the Navier-Stokes equations, we denote the numerical approximations by capital letters. Assume we have the velocity field $U^i$ and $V^i$ at the $i^{th}$ time step (time $t$) and continue condition is satisfied. We find the solution at the $(i+1)^{st}$ time step (time $t + \Delta t$) by the following three step approach (**Strang, 2007**).

The nonlinear terms are treated explicitly. This circumvents the solution of a nonlinear system, but introduces a CFL condition which limits the time step by a constant time the special resolution.

$$\begin{cases} \frac{U^*-U^i}{\Delta t} = -\left(\left(U^i\right)^2\right)_x - \left(U^iV^i\right)_y \\ \frac{V^*-V^i}{\Delta t} = -\left(U^iV^i\right)_x - \left(\left(V^i\right)^2\right)_y \end{cases} \tag{AII.1}$$

The viscosity terms are treated implicitly. If they were treated explicitly, we would have a time step restriction proportional to the special discretization squared. We have no such limitation for the implicit treatment. The price to pay is two linear systems to be solved in each time step.

$$\begin{cases} \frac{U^{**}-U^*}{\Delta t} = \frac{1}{Re}\left(U^{**}_{xx} + U^{**}_{yy}\right) \\ \frac{V^{**}-V^*}{\Delta t} = \frac{1}{Re}\left(V^{**}_{xx} + V^{**}_{yy}\right) \end{cases} \tag{AII.2}$$

We correct the intermediate velocity field $(U^{**}, V^{**})$ by the gradient of a pressure $P^{i+1}$ to enforce incompressibility.

$$\begin{cases} \frac{U^{i+1}-U^{**}}{\Delta t} = -\left(P^{i+1}\right)_x \\ \frac{V^{i+1}-V^{**}}{\Delta t} = -\left(P^{i+1}\right)_y \end{cases} \tag{AII.3}$$

The pressure is denoted $P^{i+1}$, since it is only given implicitly. It is obtained by solving a linear system. In vector notation the correction equation read as

$$\begin{cases} \frac{\mathbf{U}^{i+1}-\mathbf{U}^i}{\Delta t} = -\nabla P^{i+1} \\ -\Delta P^{i+1} = -\frac{\nabla \mathbf{U}^i}{\Delta \mathrm{t}} \end{cases} \tag{AII.4}$$

In the above numerical algorithm for fluid flow, the compact finite difference method will be used for the spatial discretization. Given the values of a function on a set of nodes the finite difference approximation to the derivative of the function is expressed as linear combination of the given function values. For simplicity consider a uniformly spaced mesh where the nodes are indexed by $i$. The independent variable at the nodes is $x_i = h(i-1)$ for $1 \le i \le N$ and the function values at the nodes $f_i = f(x_i)$ are given. The finite difference approximation $f_i{}'$ to the first derivative $df/dx(x_i)$ at the node $i$ depends on the function values at nodes near $i$. The compact finite difference schemes mimic the global dependence. The generalizations are derived by writing approximation of the form (**Lele, 1992**):

$$\alpha f'_{i-1} + f'_i + \alpha f'_{i+1} = a\frac{(f_{i+1} - f_{i-1})}{2h} \tag{AII.5}$$

The relations between the coefficient $\alpha$ and $a$ are derived by matching the Taylor series coefficients of various orders.

$$\begin{cases} f'_{i-1} = \frac{df_i}{dx} - \frac{d^2 f_i}{dx^2} h + \frac{1}{2}\frac{d^3 f_i}{dx^3} h^2 - \frac{1}{6}\frac{d^4 f_i}{dx^4} h^3 + O(h^4) \\ f'_{i+1} = \frac{df_i}{dx} + \frac{d^2 f_i}{dx^2} h + \frac{1}{2}\frac{d^3 f_i}{dx^3} h^2 + \frac{1}{6}\frac{d^4 f_i}{dx^4} h^3 + O(h^4) \\ f_{i-1} = f_i - \frac{df_i}{dx} h + \frac{1}{2}\frac{d^2 f_i}{dx^2} h^2 - \frac{1}{6}\frac{d^3 f_i}{dx^3} h^3 + \frac{1}{24}\frac{d^4 f_i}{dx^4} h^4 + O(h^5) \\ f_{i+1} = f_i + \frac{df_i}{dx} h + \frac{1}{2}\frac{d^2 f_i}{dx^2} h^2 + \frac{1}{6}\frac{d^3 f_i}{dx^3} h^3 + \frac{1}{24}\frac{d^4 f_i}{dx^4} h^4 + O(h^5) \end{cases} \quad \text{(AII.6)}$$

And the constraints are

$$\begin{cases} 1 + 2\alpha = 2a \\ \alpha = \frac{a}{3} \end{cases} \quad \text{(AII.7)}$$

And it can be obtained by:

$$\begin{cases} \alpha = \frac{1}{4} \\ a = \frac{3}{4} \end{cases} \quad \text{(AII.8)}$$

The truncation error on the right-hand side for this scheme are $4/5!\,(3\alpha - 1)h^4 f^{(5)}$.

The first derivative for the non-period boundary condition at the left boundary $i = 1$ may be obtained from a relation of the form

$$f'_1 + \alpha f'_2 = \frac{1}{h}(af_1 + bf_2 + cf_3 + df_4) \quad \text{(AII.9)}$$

Coupled to the relations written for the interior nodes. Requiring to be at least fourth-order accurate constraints the coefficients to, and the coefficient can be given by.

$$\alpha = 3; a = -\frac{17}{6}; b = \frac{3}{2}; c = \frac{3}{2}; d = -\frac{1}{6} \quad \text{(AII.10)}$$

The truncation error on the right-hand side for these boundary approximations are given by $6/5!\,h^4 f^{(5)}$.

Similarly, the first derivative at the right boundary $i = N$ can be expressed as

$$f'_N + \alpha f'_{N-1} = \frac{1}{h}(af_N + bf_{N-1} + cf_{N-2} + df_{N-3}) \quad \text{(AII.11)}$$

And the coefficients are given by

$$\alpha = 3; a = \frac{17}{6}; b = -\frac{3}{2}; c = -\frac{3}{2}; d = \frac{1}{6} \quad \text{(AII.12)}$$

Analogously, the coefficient of the second-order differential compact finite difference scheme can be obtained, which are listed in Table AII.1.

Table AII.1. Coefficients of compact finite difference

| | $\alpha$ | $a$ | $b$ | $c$ | $d$ |
|---|---|---|---|---|---|
| | first-order derivative | | | | |
| mid-point | 1/4 | 3/4 | | | |
| lower-end boundary | 3 | -17/6 | 3/2 | 3/2 | -1/6 |
| upper-end boundary | 3 | 17/6 | -3/2 | -3/2 | 1/6 |
| | second-order derivative | | | | |
| mid-point | 1/10 | 6/5 | | | |
| lower-end boundary | 11 | 13 | -27 | 15 | -1 |
| upper-end boundary | 11 | 13 | -27 | 15 | -1 |

**Appendix III: Gauss-Laguerre quadrature for double integrals**

In this appendix, a universal algorithm based on the Gauss-Laguerre quadrature for treating the double integral is developed to obtain easily and quickly pre-exponential factor of the average kernel. Consider the single integral $I$ with the $(n+1)$-point Gauss-Laguerre quadrature rule given by:

$$I = \int_0^\infty e^{-x} f(x)dx = \sum_{j=1}^{n+1} A_i f(x_i) + R_{n+1}(f) \quad \text{(AIII.1)}$$

where the abscissas $x_i$ are the zeros of the Laguerre polynomials $L_{n+1}(x)$, and $L_{n+1}(x)$ is defined as

$$L_{n+1}(x) = \frac{e^x}{(n+1)!} \frac{d^{n+1}(x^{n+1}e^{-x})}{dx^{n+1}} \quad \text{(AIII.2)}$$

and $A_i$ are the corresponding weights or Christoffel numbers, and $A_i$ can be expressed as

$$A_i = \int_{-1}^{1} \prod_{i=0, i\neq k}^{n+1} \frac{x-x_i}{x_k-x_i} e^{-x} dx \quad \text{(AIII.3)}$$

On the assumption that the definition of $f$ may be continued into the complex $z$-plane, where $z = x + iy$. Then the remainder, or truncation error, of the quadrature rule is a contour integral. The contour $C_z$ encloses the interval $-1 \leq Re(z) \leq 1$ and is such that the function $f$ is analytic on and within $C_z$. Then the remainder is given by (Donaldson and Elliott, 1972):

$$R_{n+1}(f) = \frac{1}{2\pi i} \oint k_{n+1}(z) f(z) dz \quad \text{(AIII.4)}$$

in which the function $k_{n+1}(z)$ is given by:

$$k_{n+1}(z) = \frac{\Pi_{n+1}(z)}{L_{n+1}(z)} \quad \text{(AIII.5)}$$

where

$$\Pi_{n+1}(z) = \int_0^\infty \frac{e^{-t} L_{n+1}(t)}{z-t} dt \quad \text{(AIII.6)}$$

Next, consider the double integral $II$ with $(n+1) \times (n+1)$-point Gauss-Laguerre quadrature rule given by:

$$II = \int_0^\infty \int_0^\infty e^{-x-x_1} f(x, x_1)\, dx_1 dx$$

$$= \sum_{i=1}^{n+1} \sum_{j=1}^{n+1} A_i A_j f(x_i, x_j) + R_{n+1,x_1} + R_{n+1,x} + O(R_{n+1,x}) \quad \text{(AIII.7)}$$

in which, the remainders are given by:

$$R_{n+1,x_1} = \frac{1}{2\pi i} \oint k_{n+1}(z) f(x, z) dz \quad \text{(AIII.8a)}$$

$$R_{n+1,x} = \frac{1}{2\pi i} \oint k_{n+1}(z) f(z, x_1) dz \quad \text{(AIII.8b)}$$

$$O(R_{n+1,x}) = -\frac{1}{2\pi i} \oint k_{n+1}(z) R_{n+1,x_1}(z) dz \quad \text{(AIII.8c)}$$

In the above, it is assumed that the remainder $R_{n+1,x_1}$ may be continued into the complex $z$-plane. In any case, the last term $O(R_{n+1,x})$ is essentially the remainder of remainder and henceforth it is assumed to be negligible and it is replaced by zero (Elliott et al., 2011). On combing these results, the double integral can be expressed as

$$II = \int_0^{\infty} \int_0^{\infty} e^{-x-x_1} f(x, x_1) dx_1 dx = Q_{n+1,n+1} + R_{n+1,n+1} \quad \text{(AIII.9)}$$

in which $Q_{n+1,n+1}$ and $R_{n+1,n+1}$ are given as follows:

$$Q_{n+1,n+1} = \sum_{i=1}^{n+1} \sum_{j=1}^{n+1} A_j A_j f(x_i, x_j) \quad \text{(AIII.10)}$$

and

$$R_{n+1,n+1} = 2 \int_0^{\infty} R_{n+1}(f(x, x)) dx + O(R_{n+1,x}) \quad \text{(AIII.11)}$$

For Brownian coagulation in the free molecule regime (FM),

$$\beta_{FM} = (\eta_1^{-1} + \eta^{-1})^{\frac{1}{2}} \left( \eta_1^{\frac{1}{3}} + \eta^{\frac{1}{3}} \right)^2 \quad \text{(AIII.12)}$$

The change of the value $Q_{n+1,n+1}$ and its error $\epsilon_n$ as the interpolation points increases is shown in Fig.1. and the error $\epsilon_n$ is defined as

$$\epsilon_n = |Q_{n+1,n+1} - Q_{n,n}| \quad \text{(AIII.13)}$$

As the number of evaluation points increases, the value of $Q_{n+1,n+1}$ gradually tends towards a fixed value; and the error decreases sharply, there appear to a constant slope in the error plot on the logarithmic coordinate system. According to the two-point formula of the line, it means that the following equation holds

$$\frac{\ln \epsilon_m - \ln \epsilon_n}{\ln m - \ln n} = C, \ (m, n \in N) \quad \text{(AIII.14)}$$

in which $C$ is the constant slope. And it can be reorganized into

$$\frac{\epsilon_m}{\epsilon_n} = \left(\frac{m}{n}\right)^C \quad \text{(AIII.15)}$$

It can be extended to the real number field as

$$\epsilon_x = \epsilon_n \left(\frac{x}{n}\right)^C, \ (x \in R) \quad \text{(AIII.16)}$$

And the truncation error can be calculated as

$$R_{n+1,n+1} = \int_{n+1}^{\infty} \epsilon_x dx = -\epsilon_n \left(\frac{n+1}{n}\right)^C \left(\frac{n+1}{C+1}\right); \quad (C < -1, C \in R) \quad \text{(AIII.17)}$$

In this article, the maximum number of the interpolation points is $n = 360$, and the corresponding error is $\epsilon_n = 1.2344e - 4$, the constant slope is $C = -1.5209$, and $Q_{n+1,n+1} = 6.9032$ and its remainder $R_{n+1,n+1} = 0.0852$.

Table 1. the comparison between temporal and spatial mixing layer

| Aspect | Temporal Mixing Layer | Spatial Mixing Layer |
|---|---|---|
| Configuration | Two infinite fluid layers initially at rest relative to each other | Two parallel streams with different velocities flowing past each other |
| Evolution | In time at a fixed location | In space as flow moves downstream |
| Frame of Reference | Moving with average velocity of layers | Laboratory-fixed |
| Boundary Conditions | Often periodic in streamwise direction | Inflow and outflow |
| Growth | Layer thickness grows with time | Layer thickness grows with downstream distance |
| Entrainment | No continuous entrainment of fresh fluid | Continuous entrainment from freestreams |
| Self-Similarity | Achieved in scaled time coordinates | Achieved in scaled spatial coordinates |
| Vortex Dynamics | Evolve and interact over time | Continuously generated and convected downstream |
| Energy Input | Fixed amount at start | Continuous input from freestreams |
| Computational Complexity | Simpler due to periodic conditions | Requires careful inflow/outflow treatment |
| Experimental Realization | Challenging; often studied numerically | Easily created in wind tunnels/water channels |
| Reynolds Number Evolution | Increases with time | Increases with downstream distance |
| Turbulence Development | Uniform across layer | Progresses downstream |
| Applicability | Fundamental studies, transient phenomena | Practical engineering flows |
| Scaling Laws | Growth rate scales with time | Growth rate scales with downstream distance |
| Initial Conditions | Initial perturbations crucial | Inflow conditions and splitter plate important |

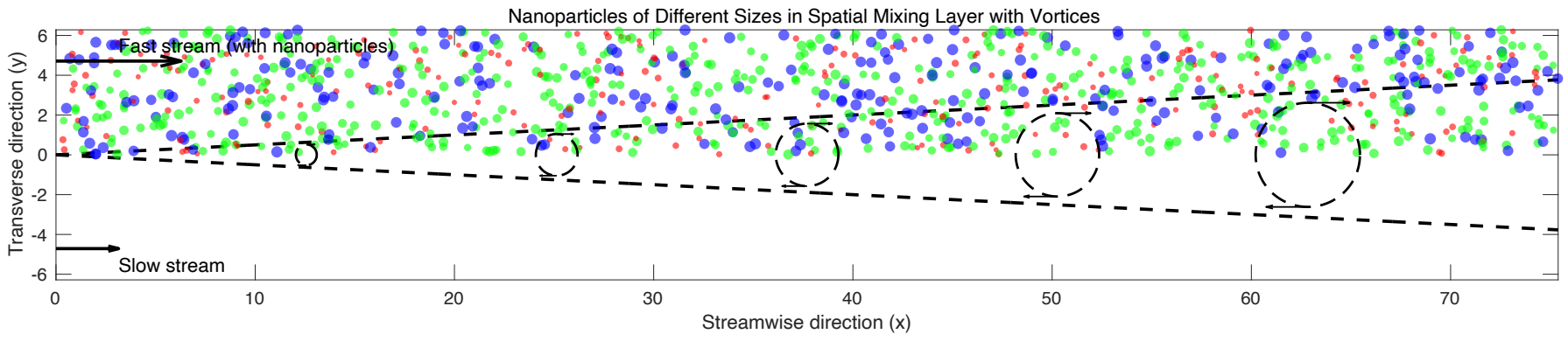

Figure 1. the configuration of nanoparticles in the spatial mixing layer

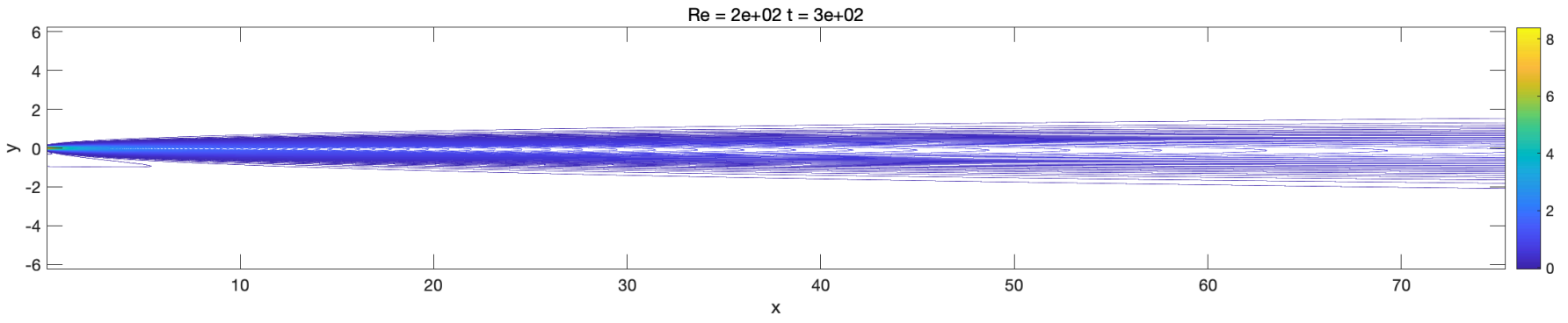

a)

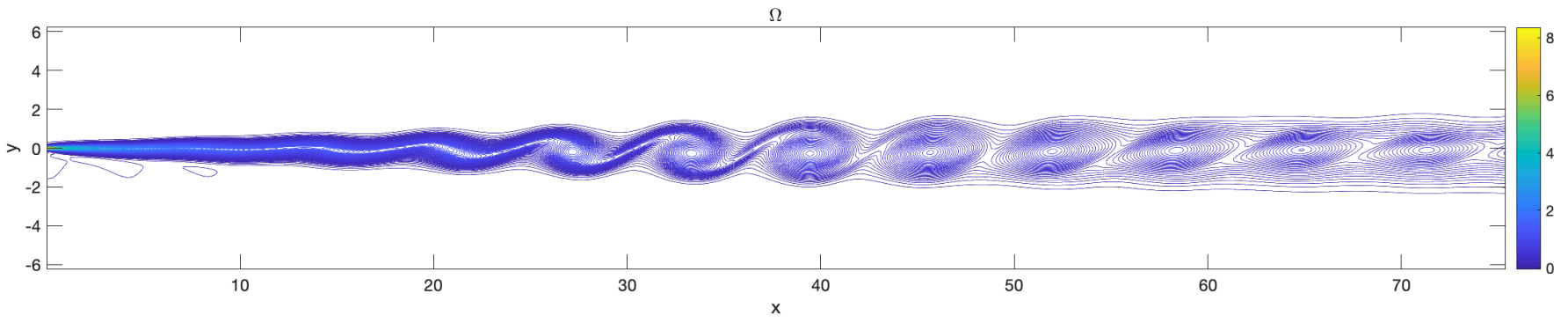

b)

Figure 2. Snapshots of vorticity of spatial mixing layer at Reynolds number 200. a) the unperturbed mixing layer; b) the forced mixing layer.

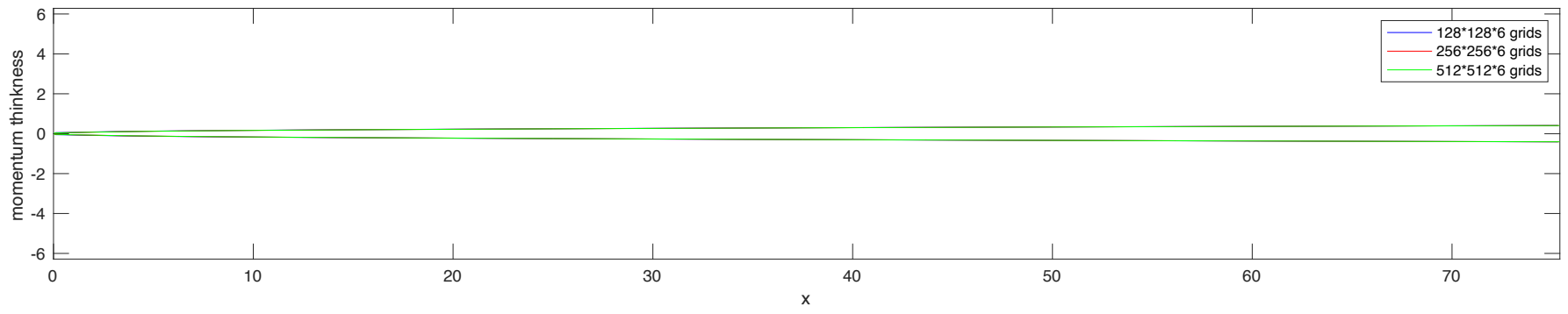

a)

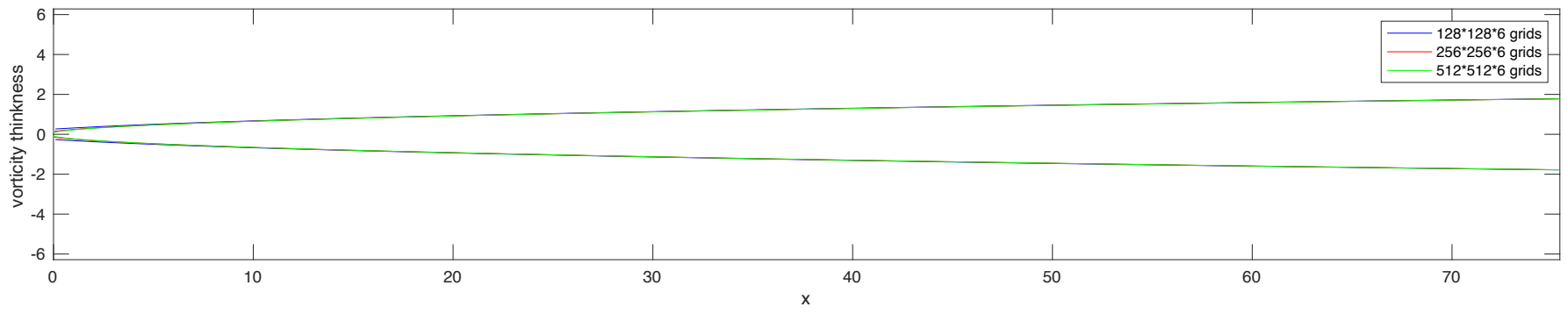

b)

Figure 3. The effect of resolution on the momentum and vorticity thickness for the unperturbed mixing layer at Reynolds number 200. a) the streamwise growth of momentum thickness; b) the streamwise growth of vorticity thickness

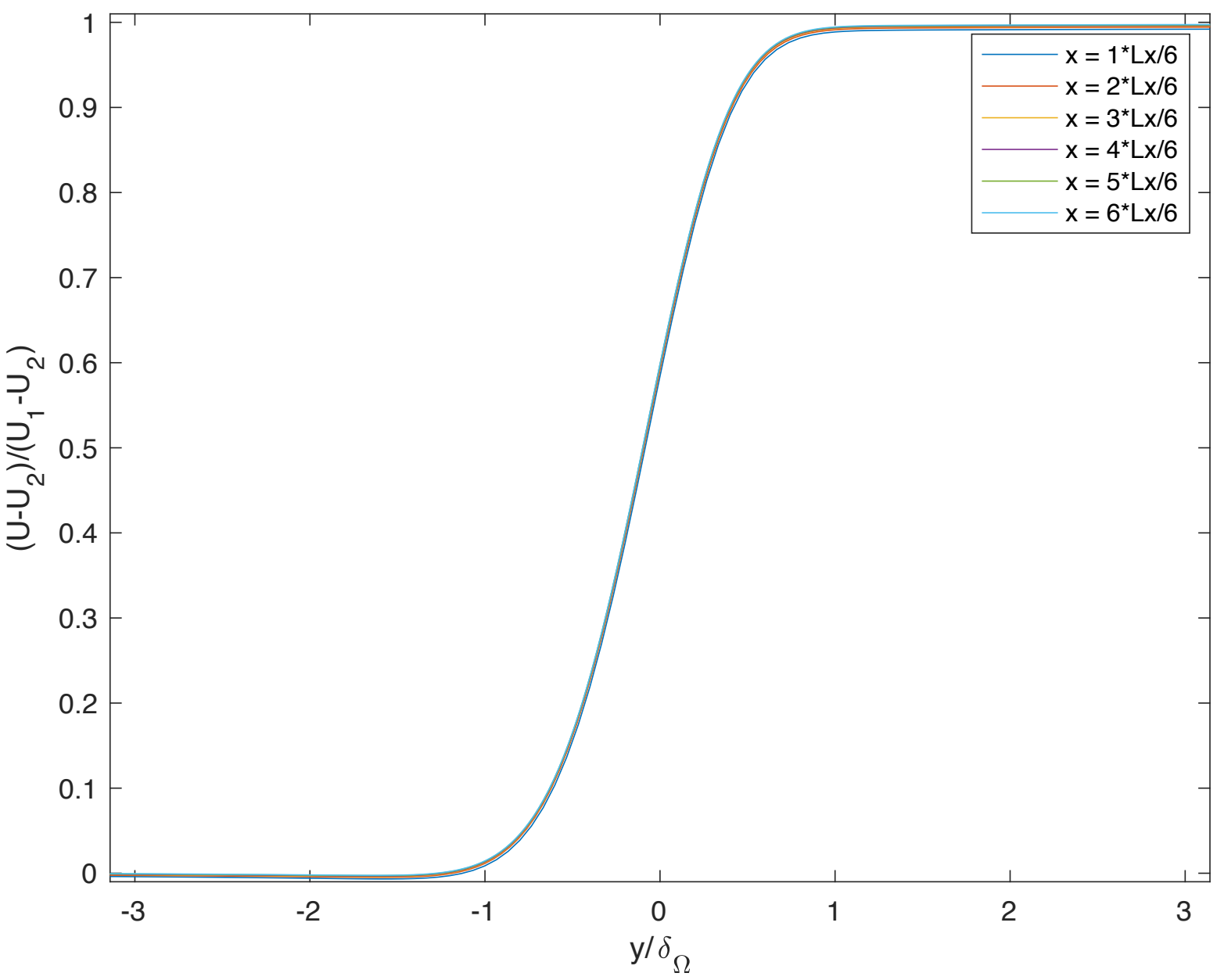

(a)

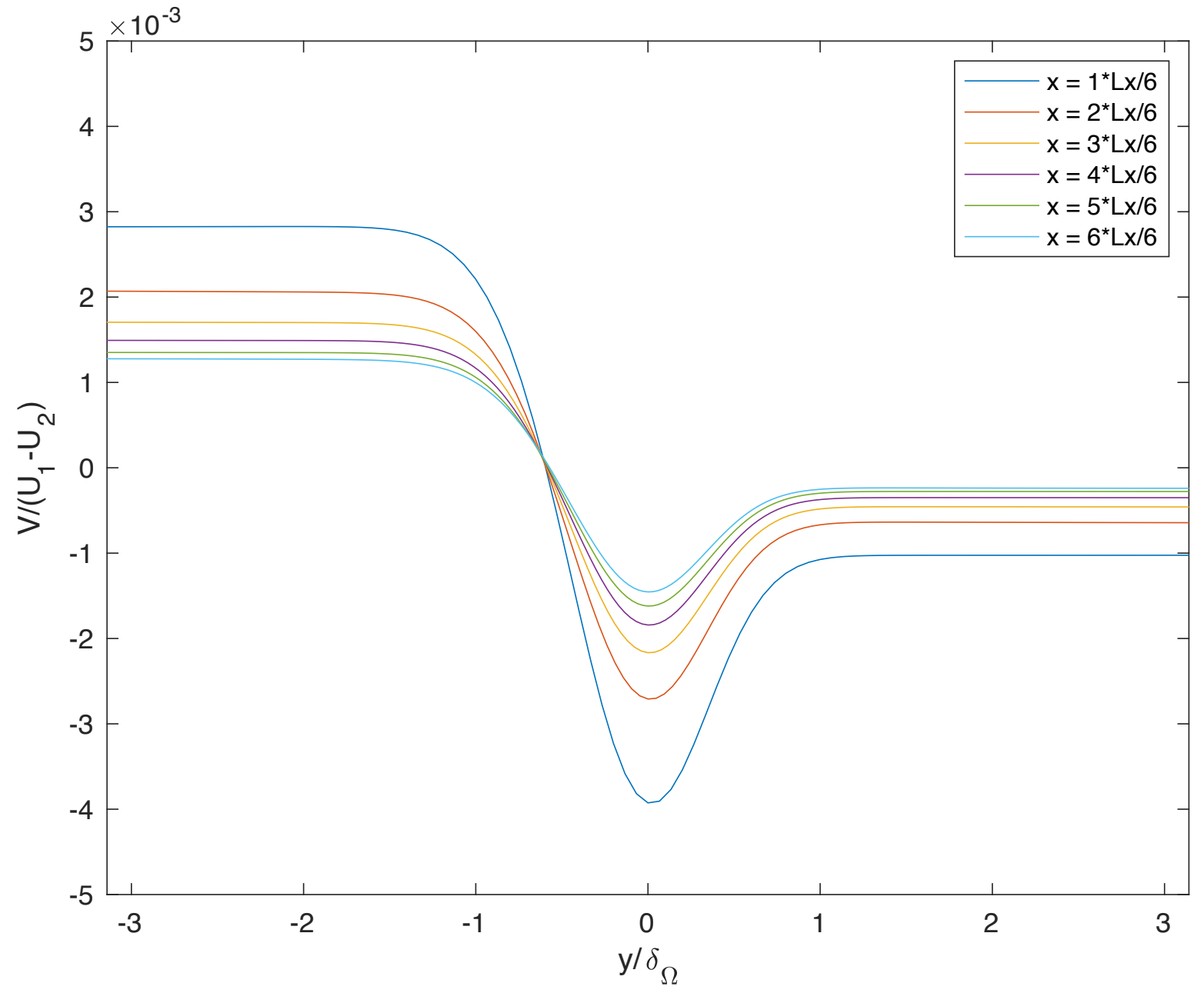

(b)

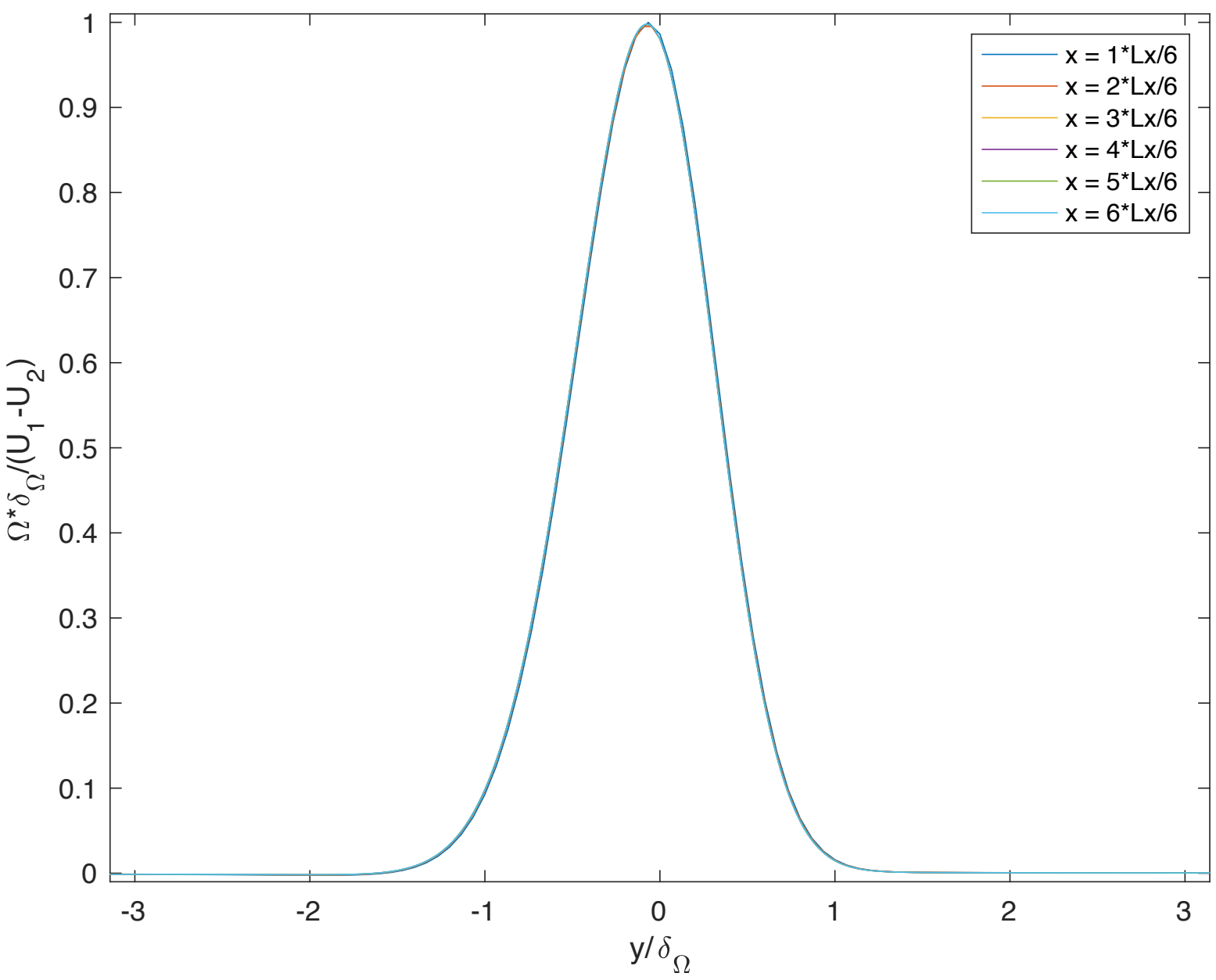

(c)

Figure 4. The similarity of the distribution of velocity and vorticity at different streamwise location. a) mean field statistic for streamwise velocity; b) mean field statistic for transverse velocity; c) mean field statistic for vorticity.

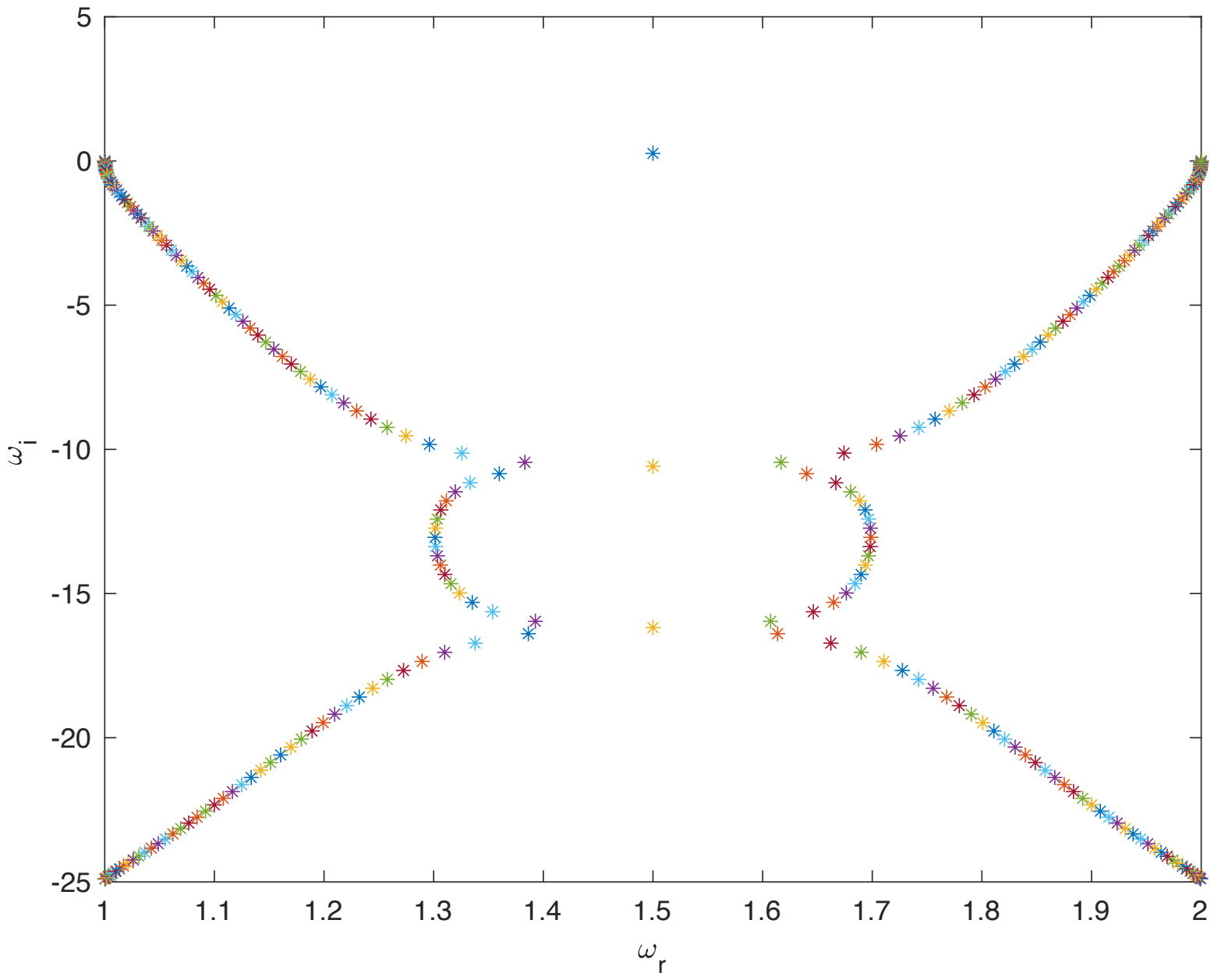

(a)

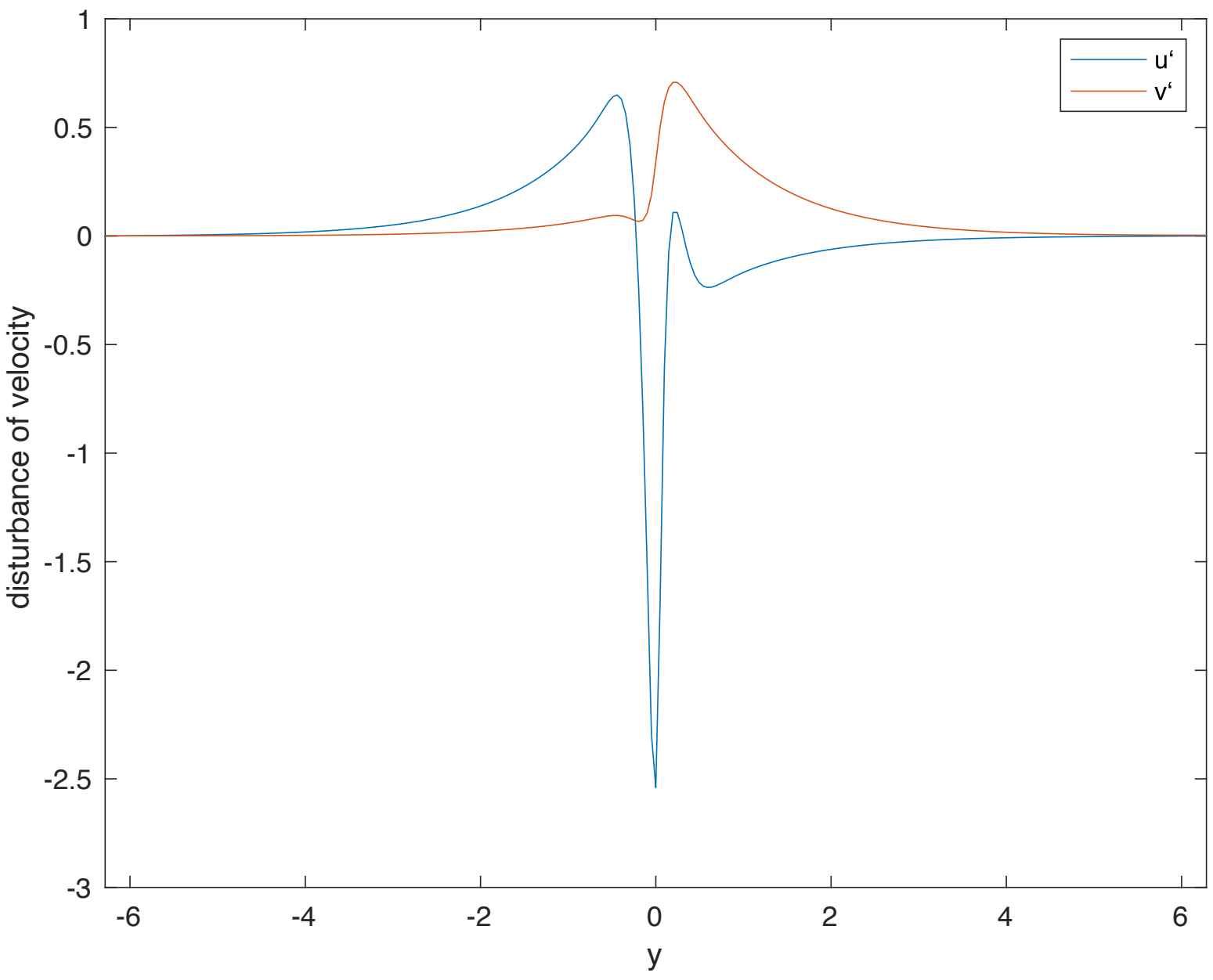

(b)

Figure 5. Distribution of eigenvalue and eigenfunctions based on linear stability theory for parallel mixing flow. a) the distribution of eigenvalue b) the distribution of eigenfunctions with maximum imaginary part of eigenvalue is $\omega_i = 0.2647$

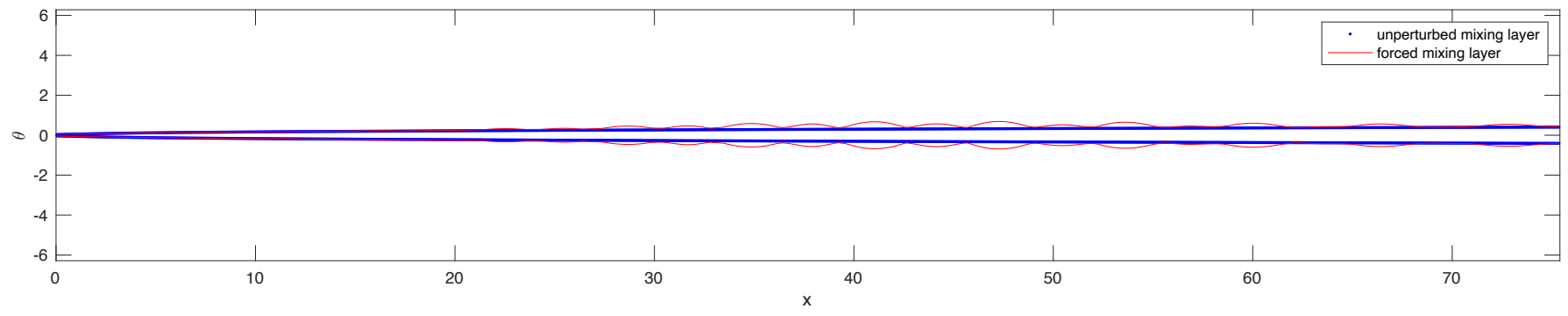

(a)

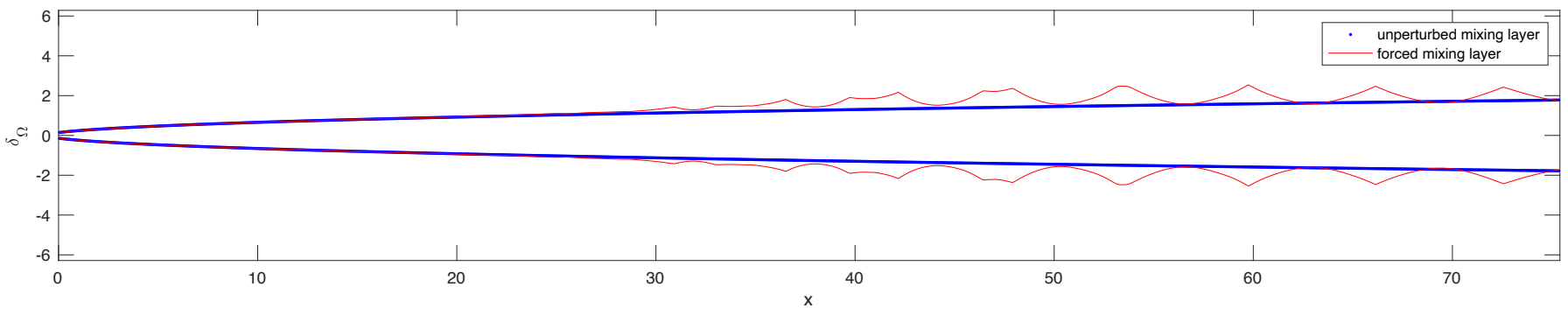

(b)

Figure 6. The streamwise development of the momentum and vorticity thickness in the forced mixing layer at Reynolds number 200 and $1536 \times 256$ grids. a) the streamwise growth of momentum thickness; b) the streamwise growth of vorticity thickness

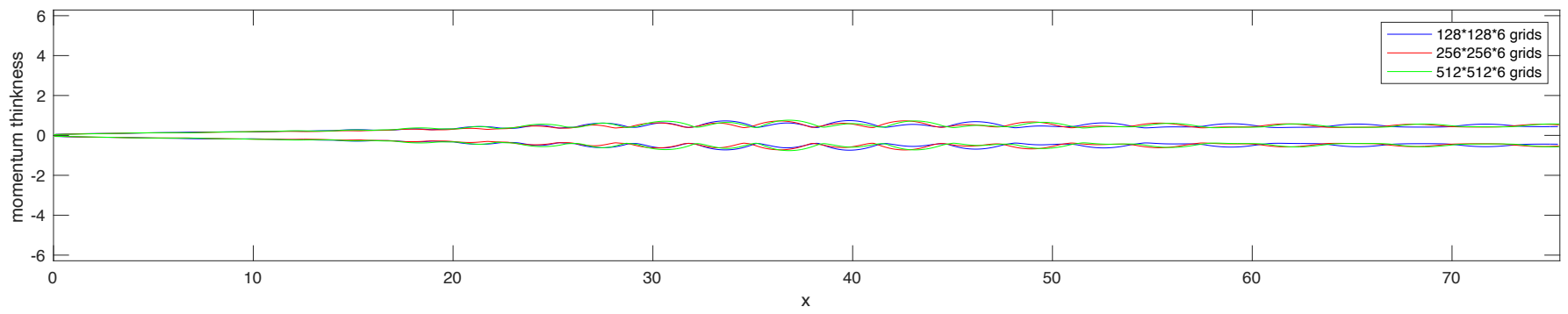

(a)

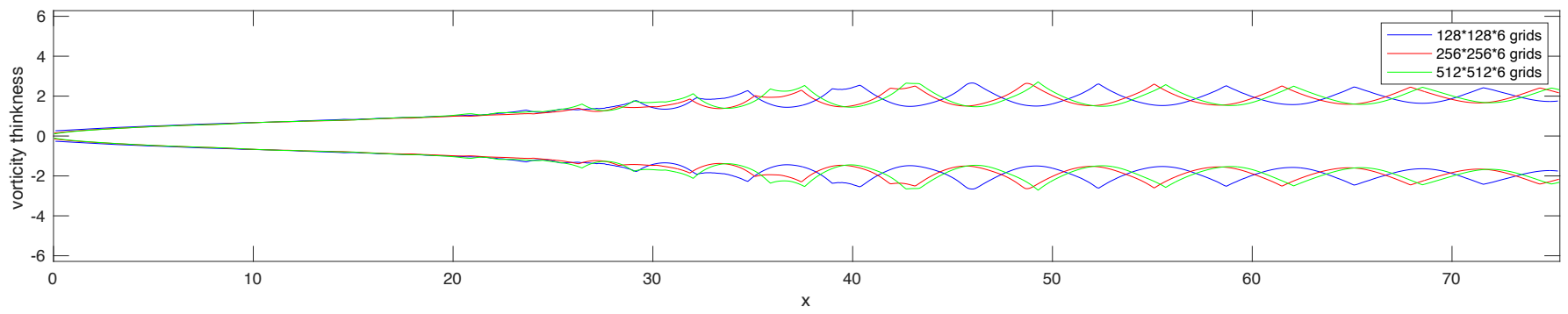

(b)

Figure 7. The effect of resolution on the momentum and vorticity thickness for the forced mixing layer at Reynolds number 200. a) the streamwise growth of momentum thickness; b) the streamwise growth of vorticity thickness

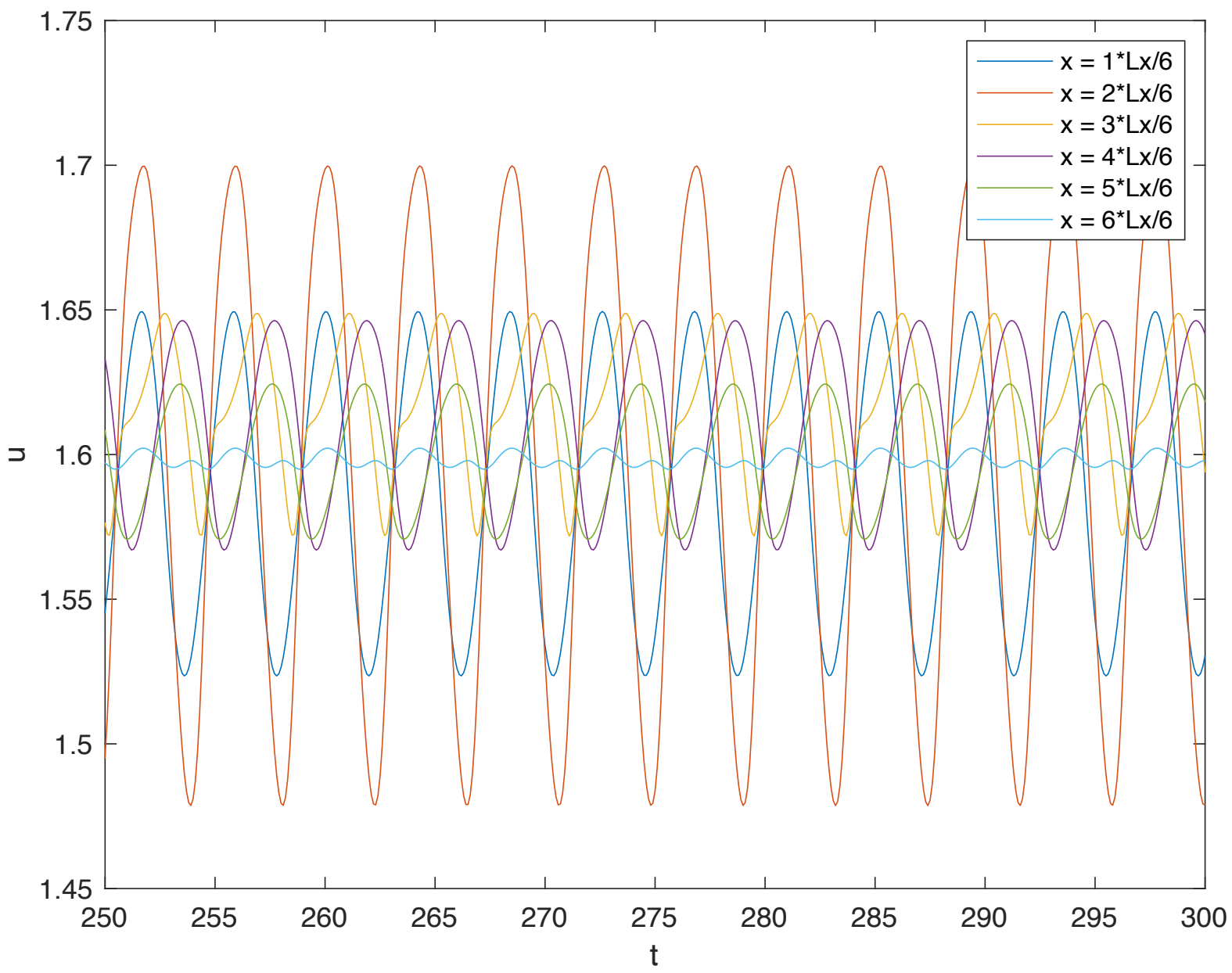

a)

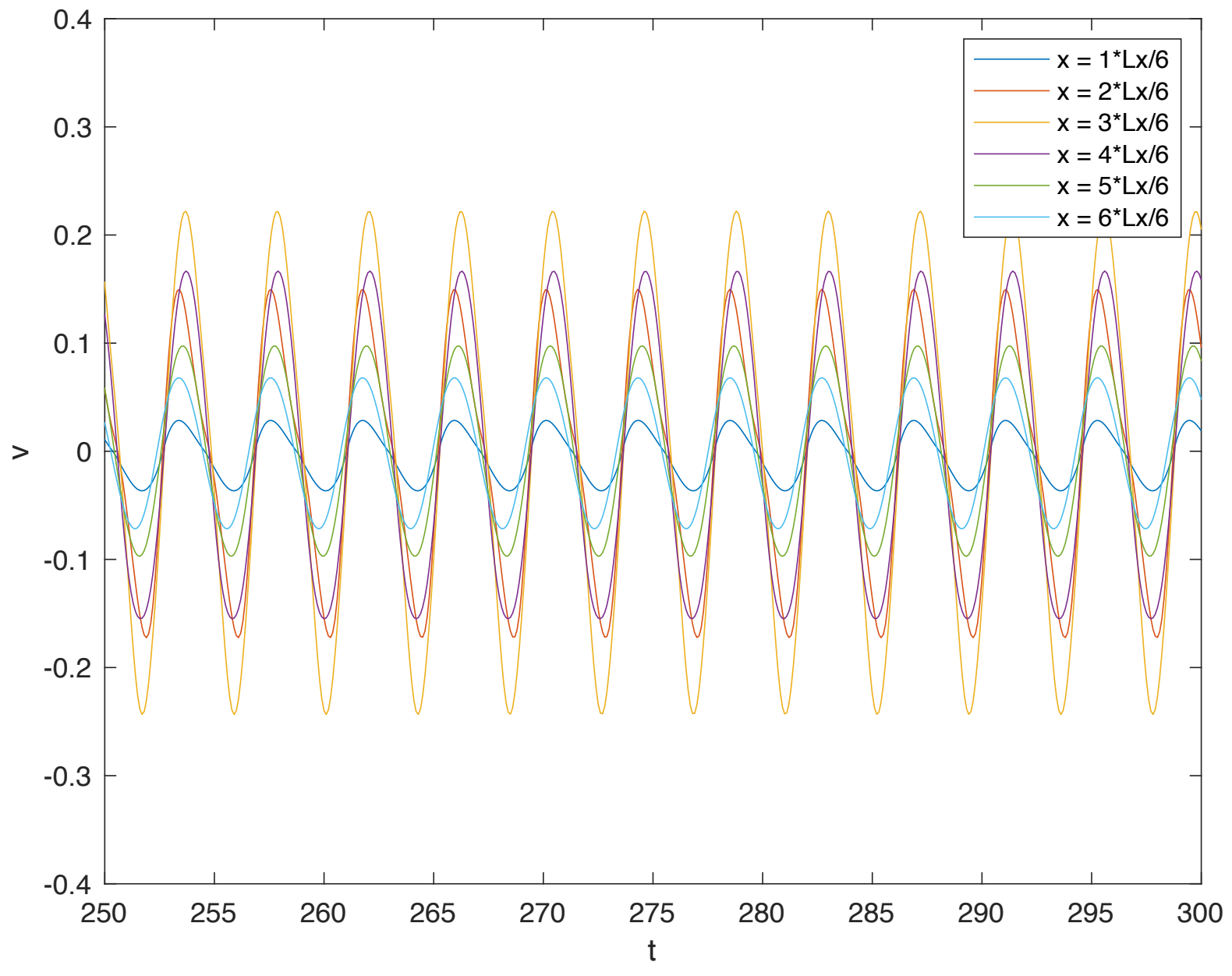

b)

Figure 8. Velocity time histories at different central streamwise location. a) for the $u$ component; b) for the $v$ component

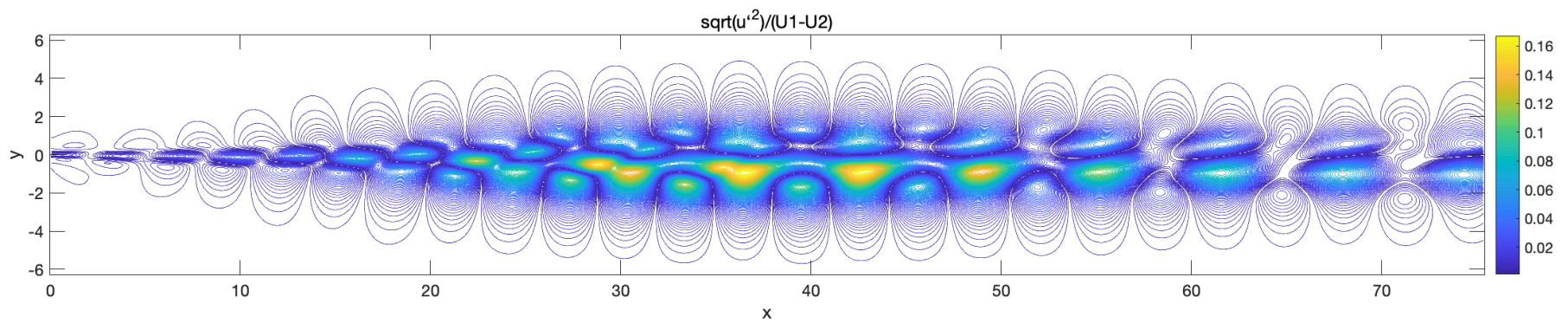

a)

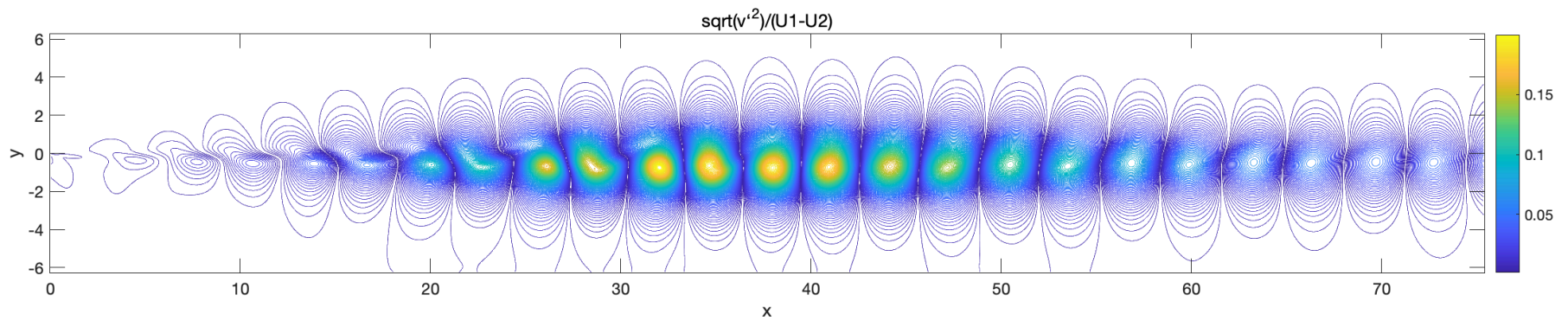

b)

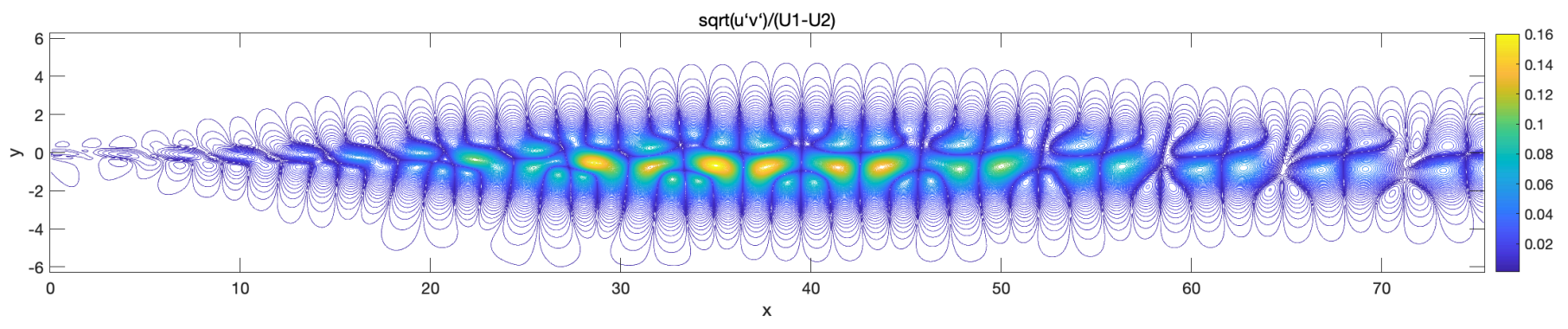

c)

Figure 9. The spatial distribution of turbulence intensity at $Re = 200$ and $t = 300$, a) for $\sqrt{u'^2}/(U_1 - U_2)$ ; b) for $\sqrt{v'^2}/(U_1 - U_2)$; c) for $\sqrt{u'v'}/(U_1 - U_2)$

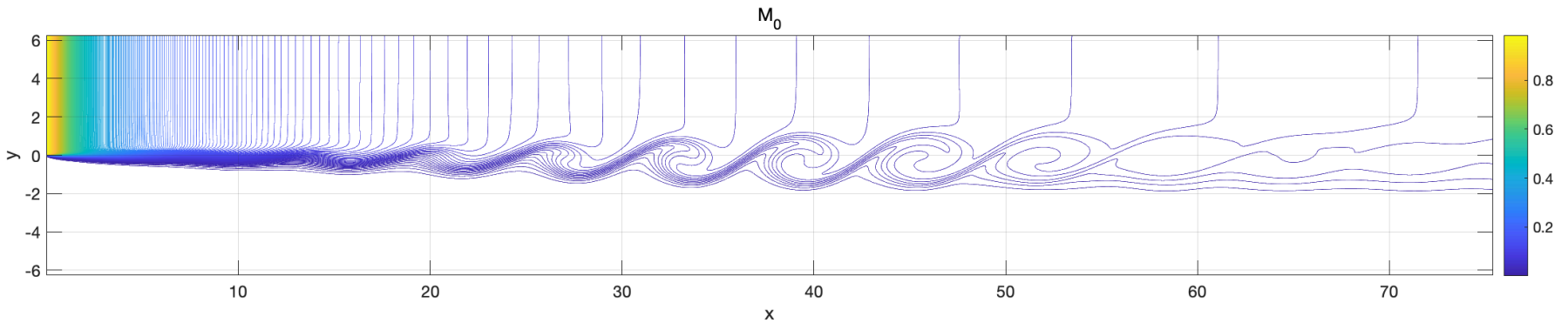

(a)

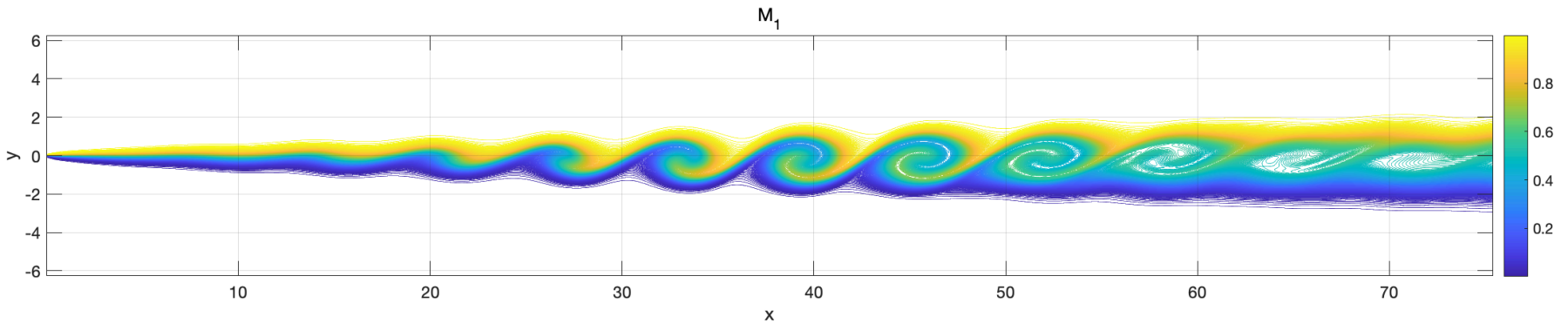

(b)

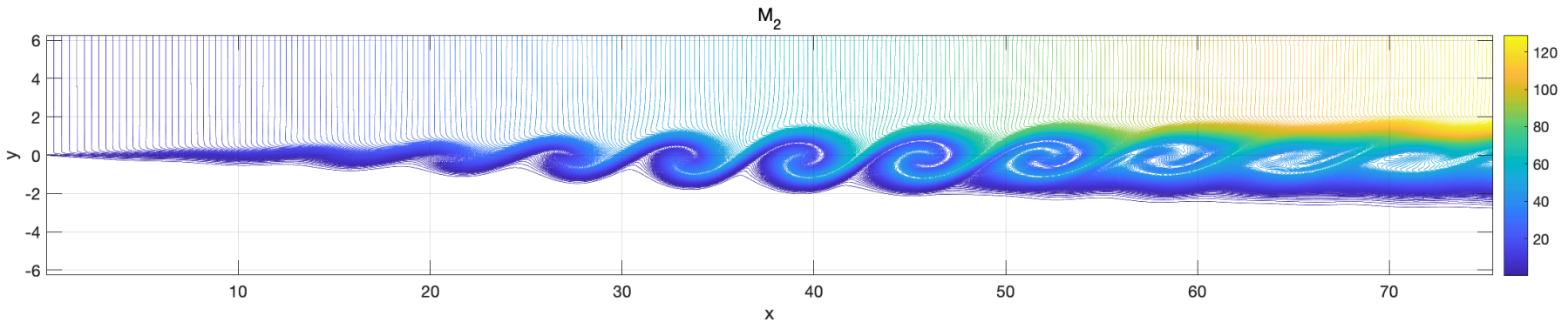

(c)

Figure 10. the distribution of particle moment at $Re = 200, Sc = 1, Da = 1$ and $t = 300$, a) the distribution of total particle number density ($M_0$); b) the distribution of total particle volume concentration ($M_1$); c) the distribution of dispersity of particle size distribution ($M_2$)

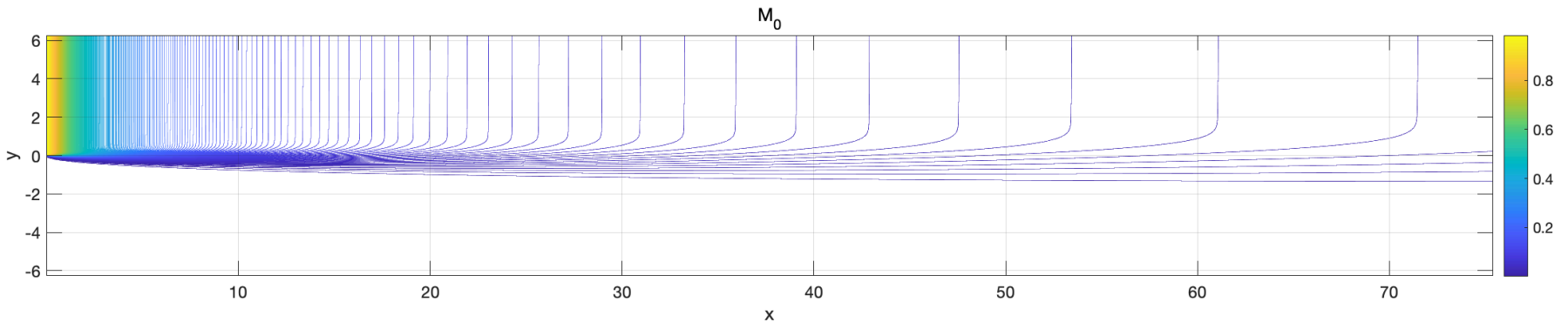

(a)

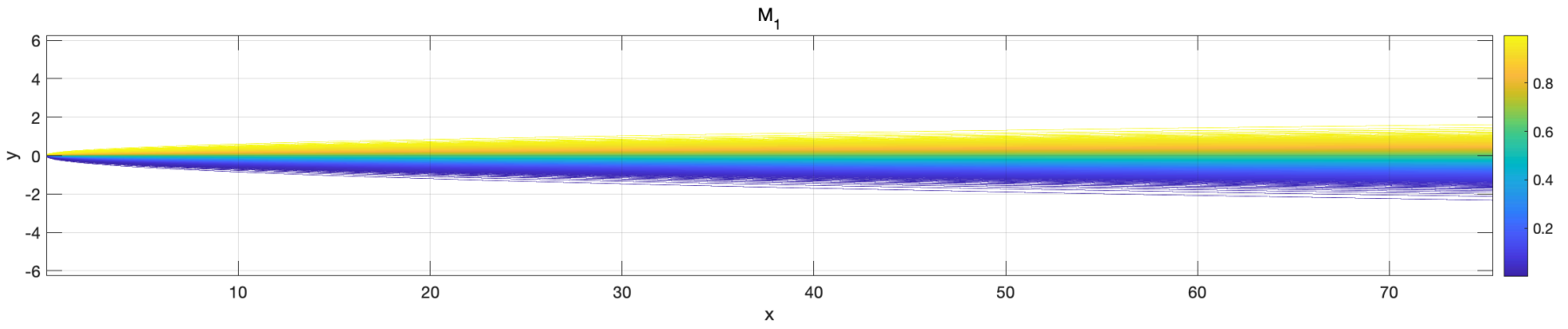

(b)

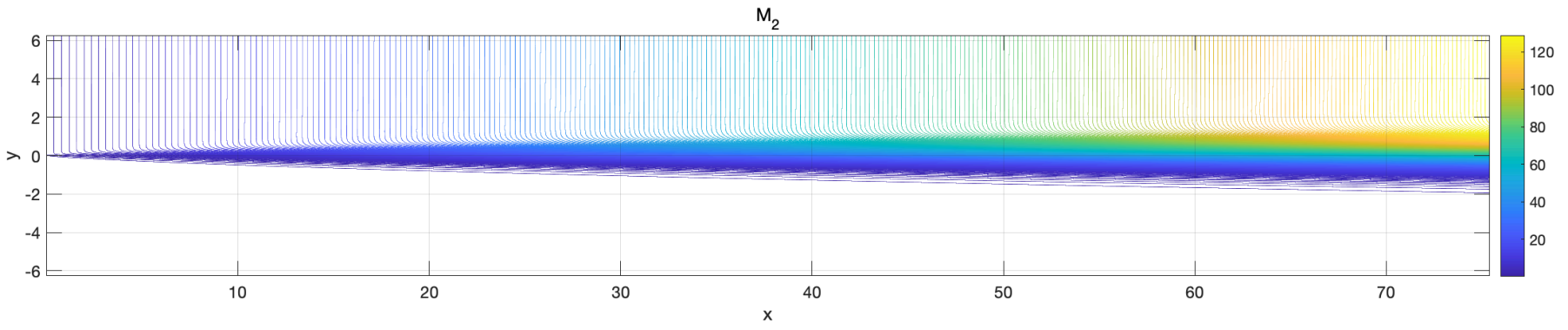

(c)

Figure 11. the distribution of mean particle moments at $Re = 200, Sc = 1, Da = 1$. a) the distribution of total particle number density ($\overline{M_0}$); b) the distribution of total particle volume concentration ($\overline{M_1}$); c) the distribution of dispersity of particle size distribution ($\overline{M_2}$)

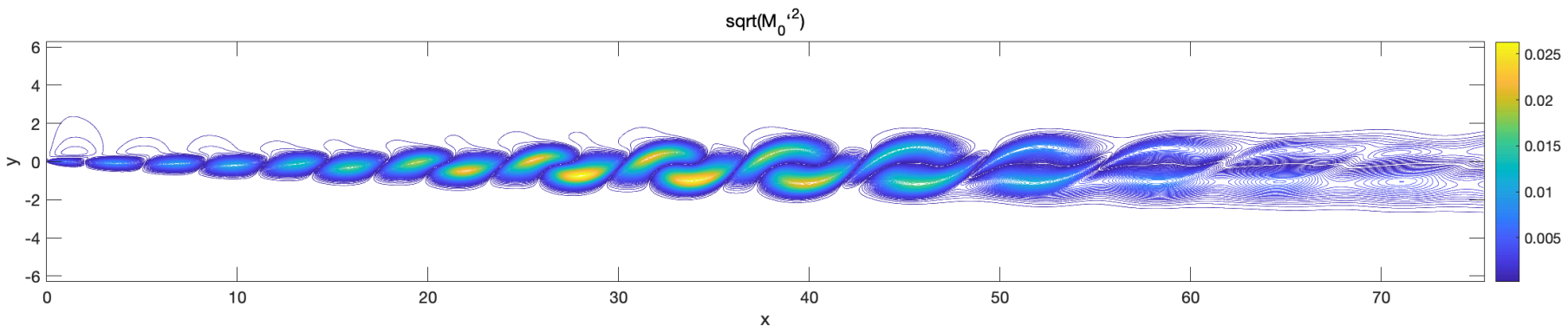

(a)

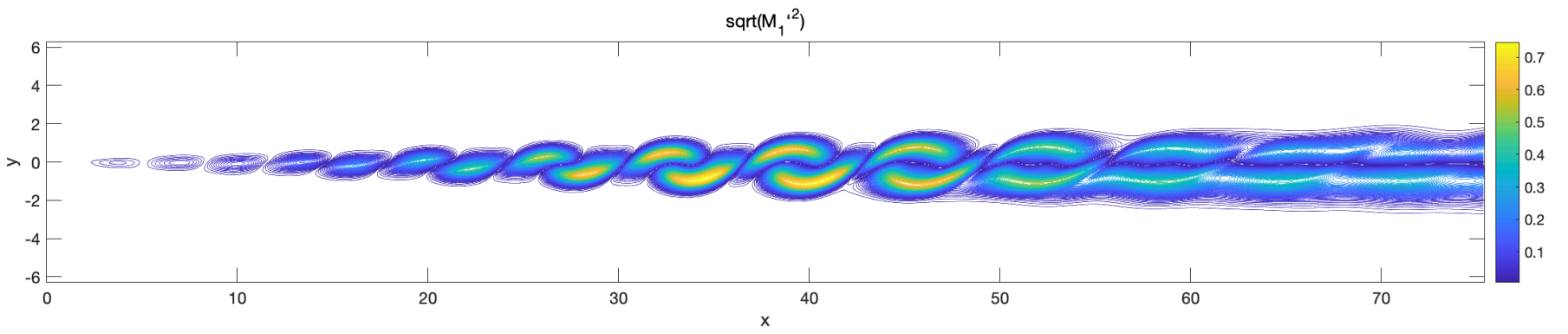

(b)

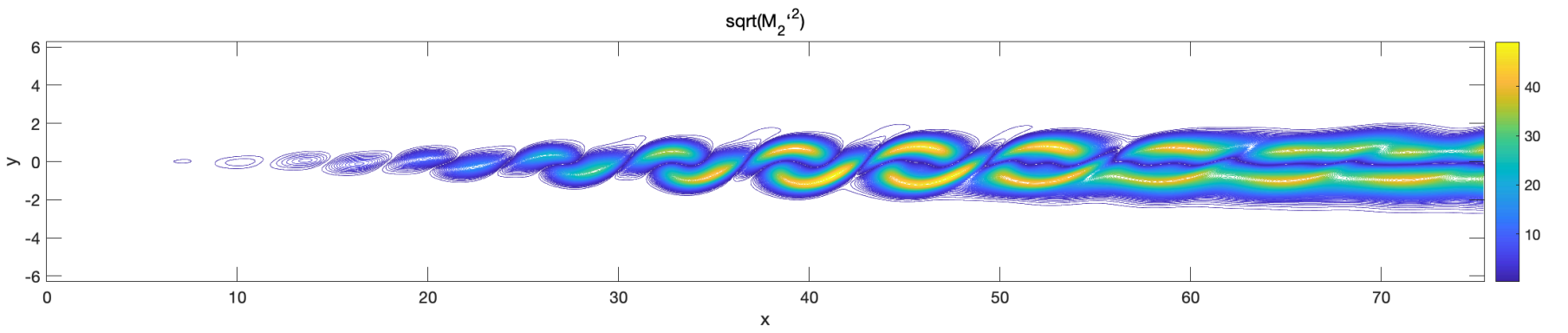

(c)

Figure 12. the distribution of perturbated particle moments at $Re = 200, Sc = 1, Da = 1$ and $t = 300$. a) the distribution of total particle number density ($\sqrt{M_0'^2}$); b) the distribution of total particle volume concentration ($\sqrt{M_1'^2}$); c) the distribution of dispersity of particle size distribution ($\sqrt{M_2'^2}$).

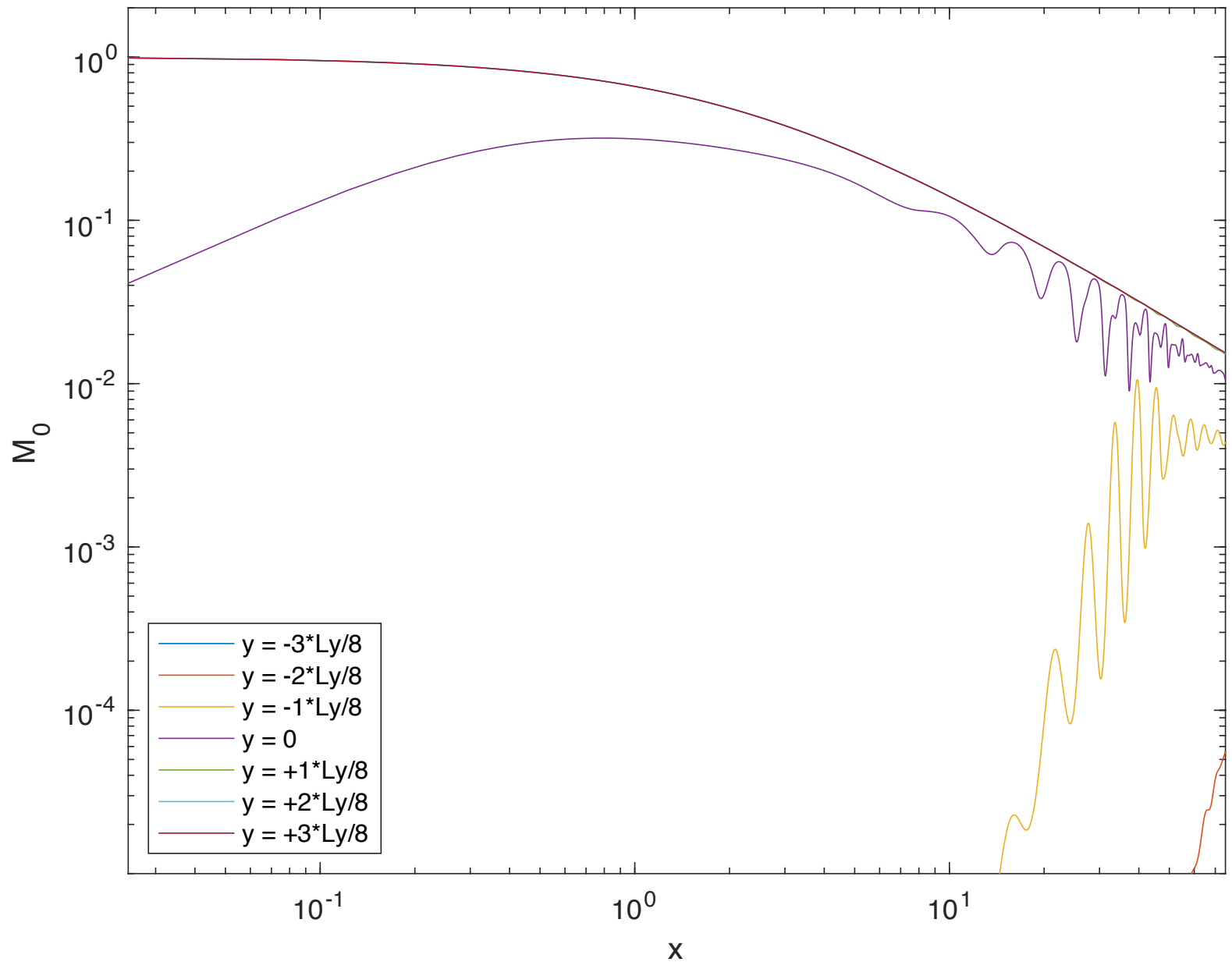

(a)

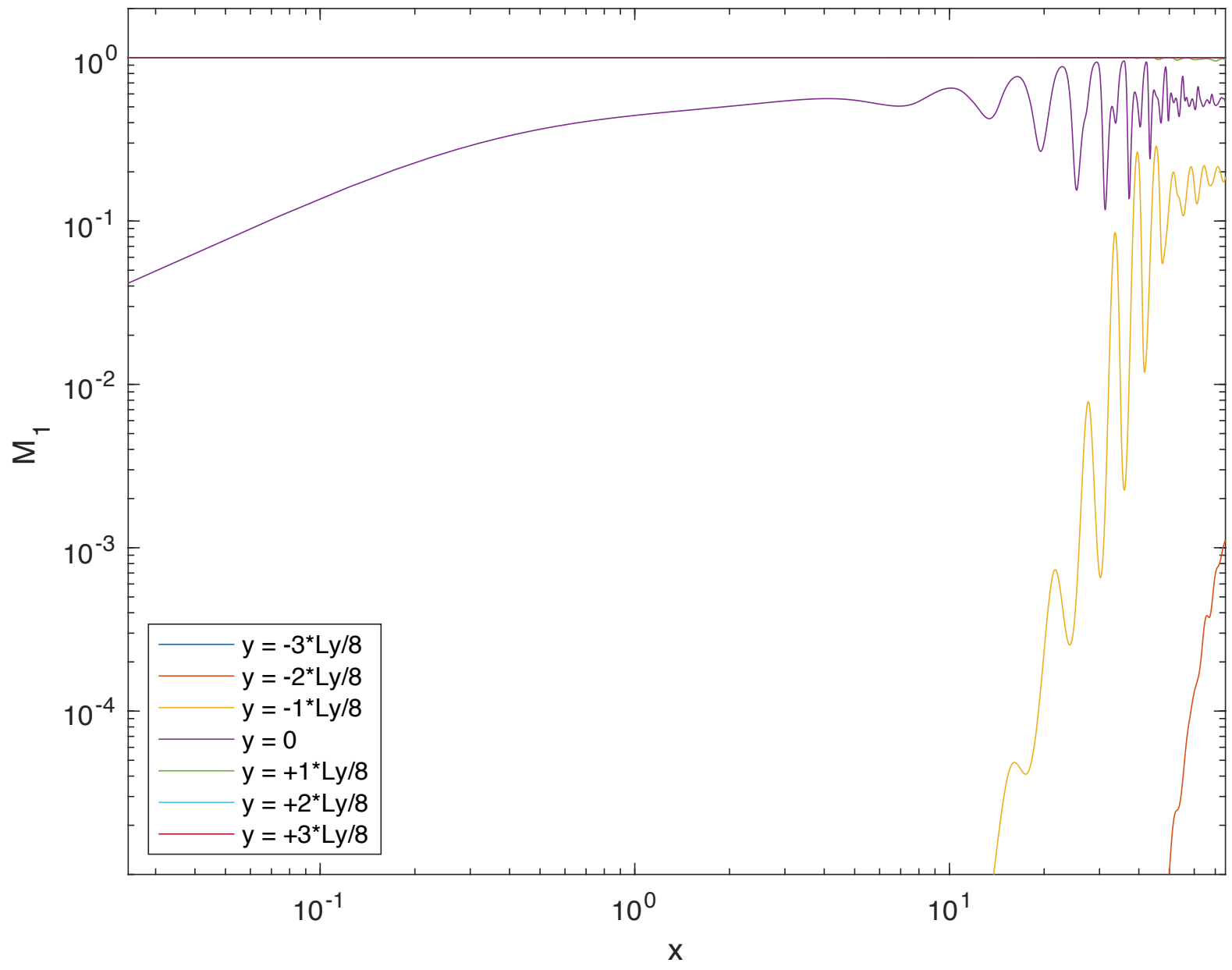

(b)

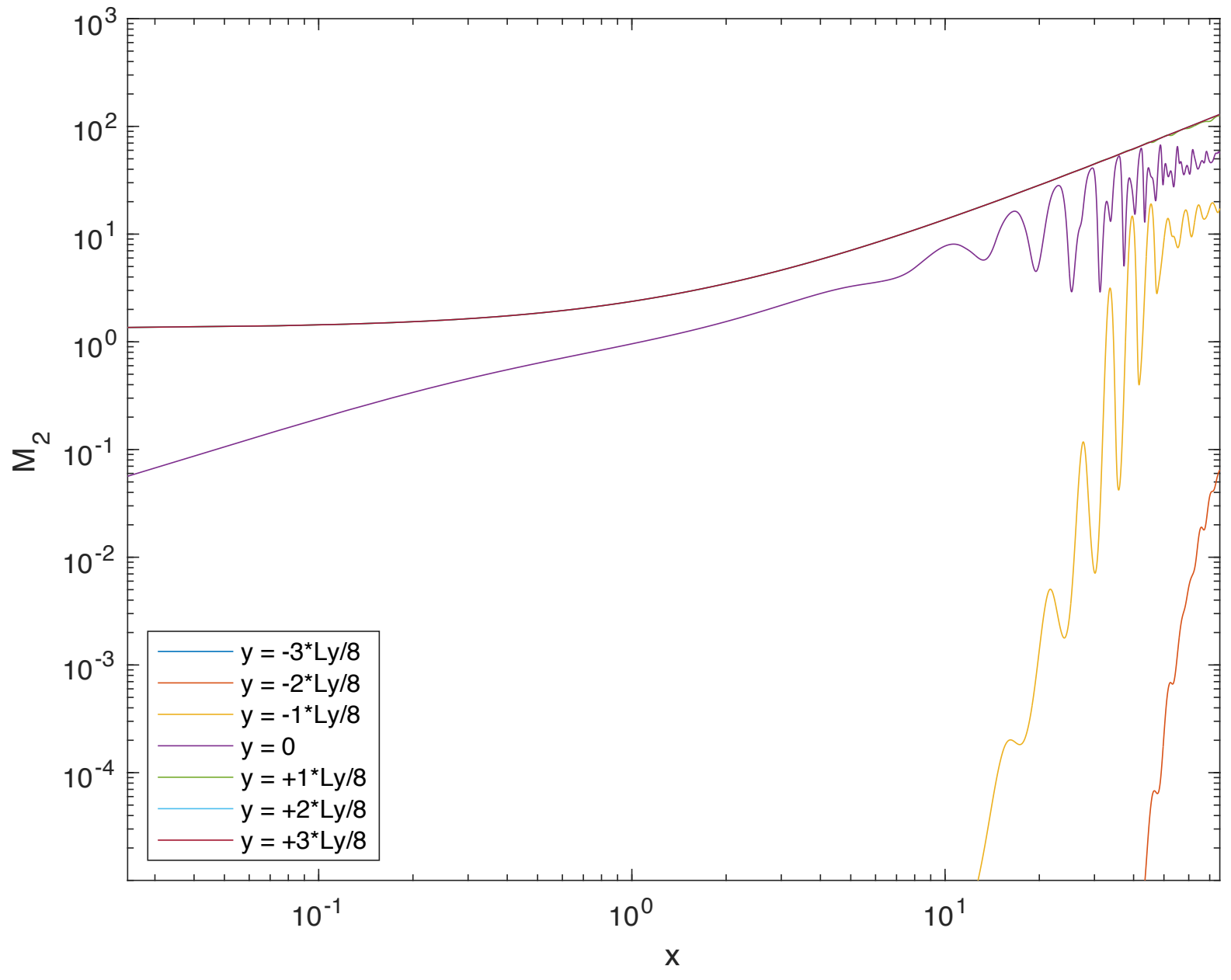

(c)

Figure 13. the evolution particle moments along the streamwise direction at $Re = 200, Sc = 1, Da = 1$ and $t = 300$, a) the distribution of total particle number density ($M_0$); b) the distribution of total particle volume concentration ($M_1$); c) the distribution of dispersity of particle size distribution ($M_2$)

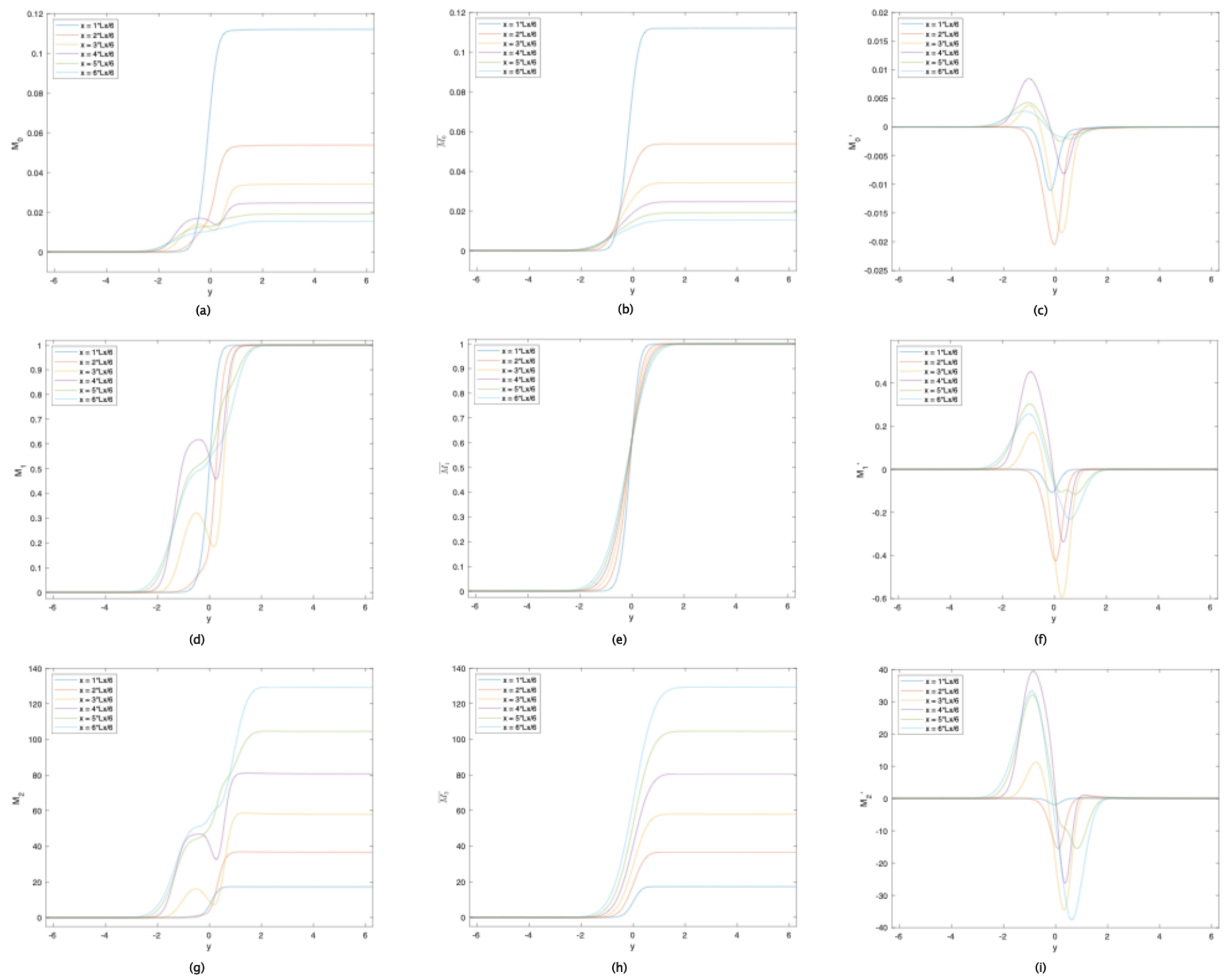

Figure 14. the evolution particle moment along the transverse direction at $Re = 200, Sc = 1, Da = 1$. a) the distribution of total particle number density ($M_0$); b) the distribution of mean particle number density ($\overline{M_0}$); c) the distribution of fluctuating particle number density (${M_0}'$); d) the distribution of total particle volume concentration ($M_1$); e) the distribution of mean particle volume concentration ($\overline{M_1}$); f) the distribution of fluctuating particle volume concentration (${M_1}'$); g) the distribution of dispersity of particle size distribution ($M_2$); h) the distribution of mean dispersity of particle size distribution ($\overline{M_2}$); i) the distribution of fluctuating dispersity of particle size distribution (${M_2}'$).

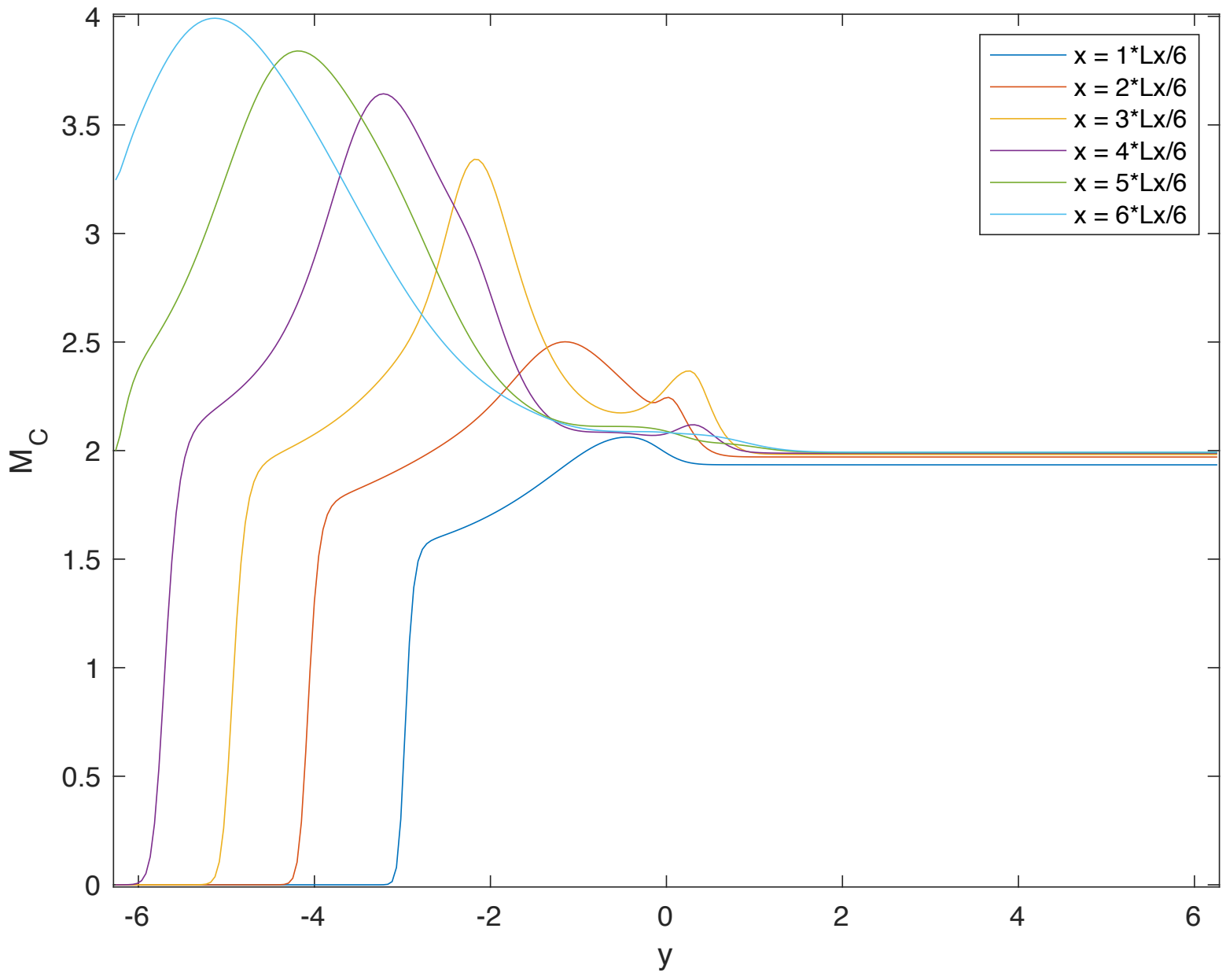

Figure 15. the distribution of particle dimensionless moment $(Re)$ at $Re = 200, Sc = 1, Da = 1$ and $t = 300$.

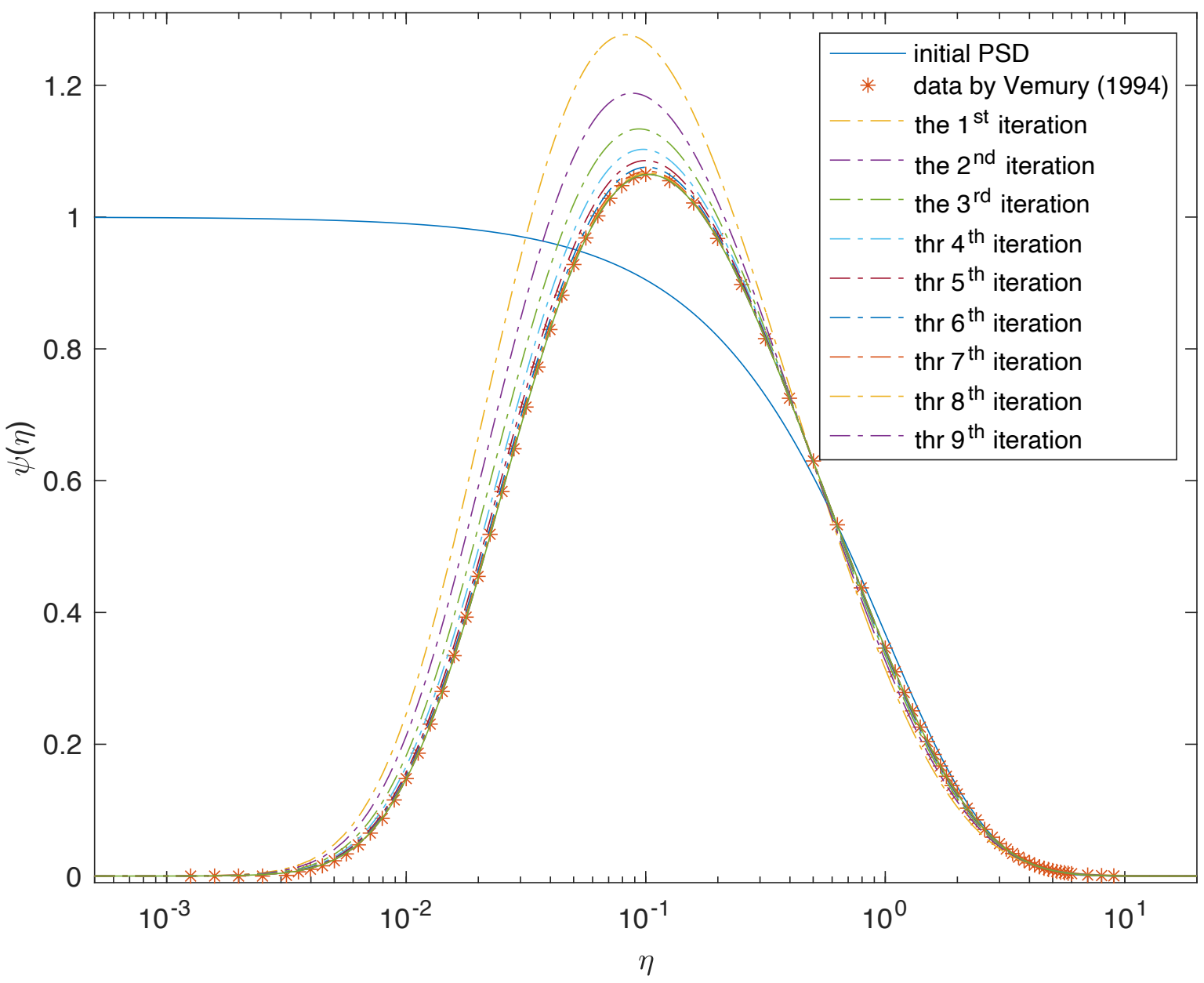

Figure 16. the corrected similarity solution for 0-dimenaional coagulating system based on AK-iDNS framework

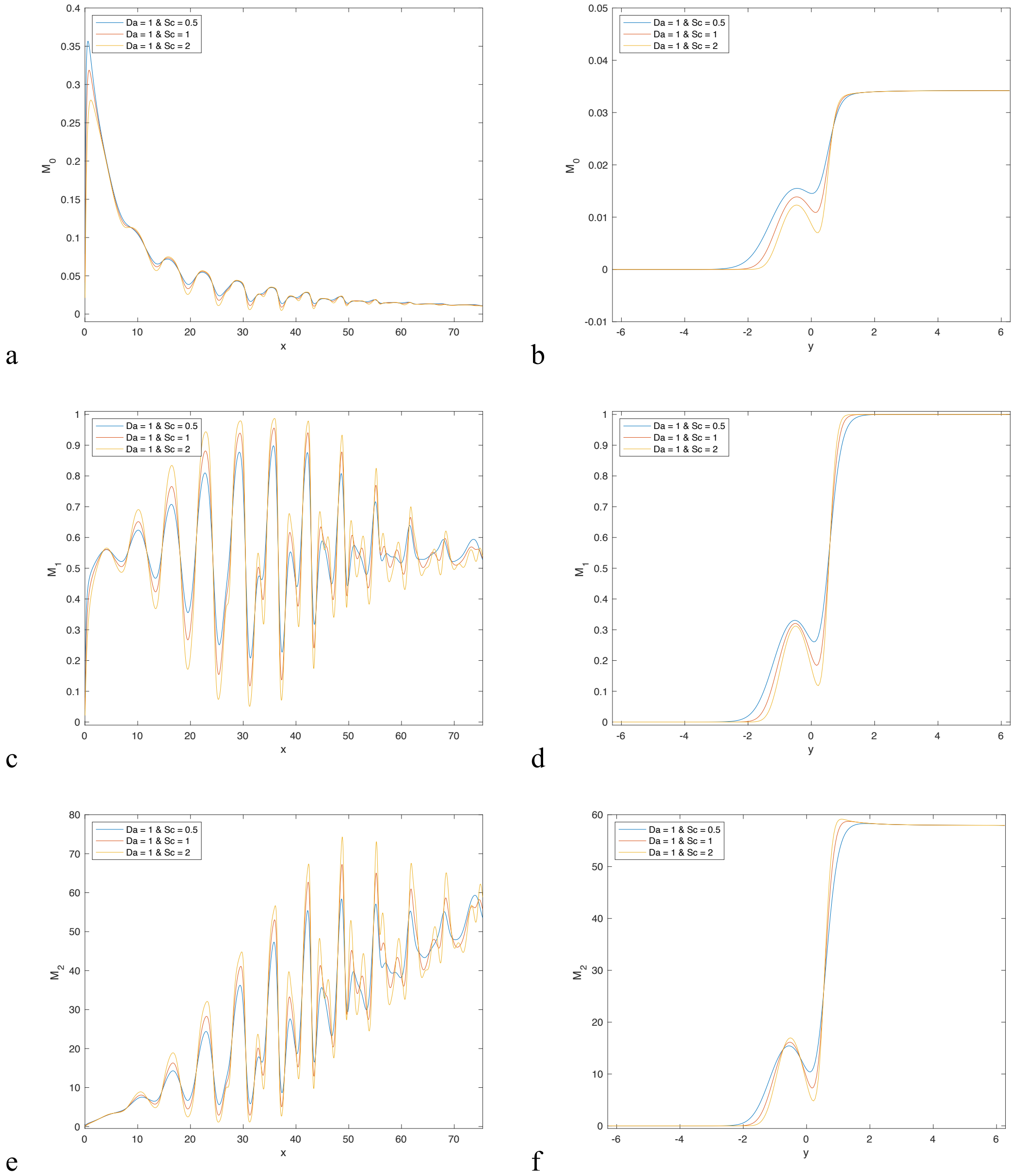

Figure 17. The effect of Schmidt number ($Sc$) on the distribution of particle moment in space under the conditions $Re$ = 200, $Sc$ = 1. a) $M_0$ along the central horizontal line; b) $M_0$ along the central vertical line; c) $M_1$ along the central horizontal line; d) $M_1$ along the central vertical line; e) $M_2$ along the central horizontal line; f) $M_2$ along the central vertical line.

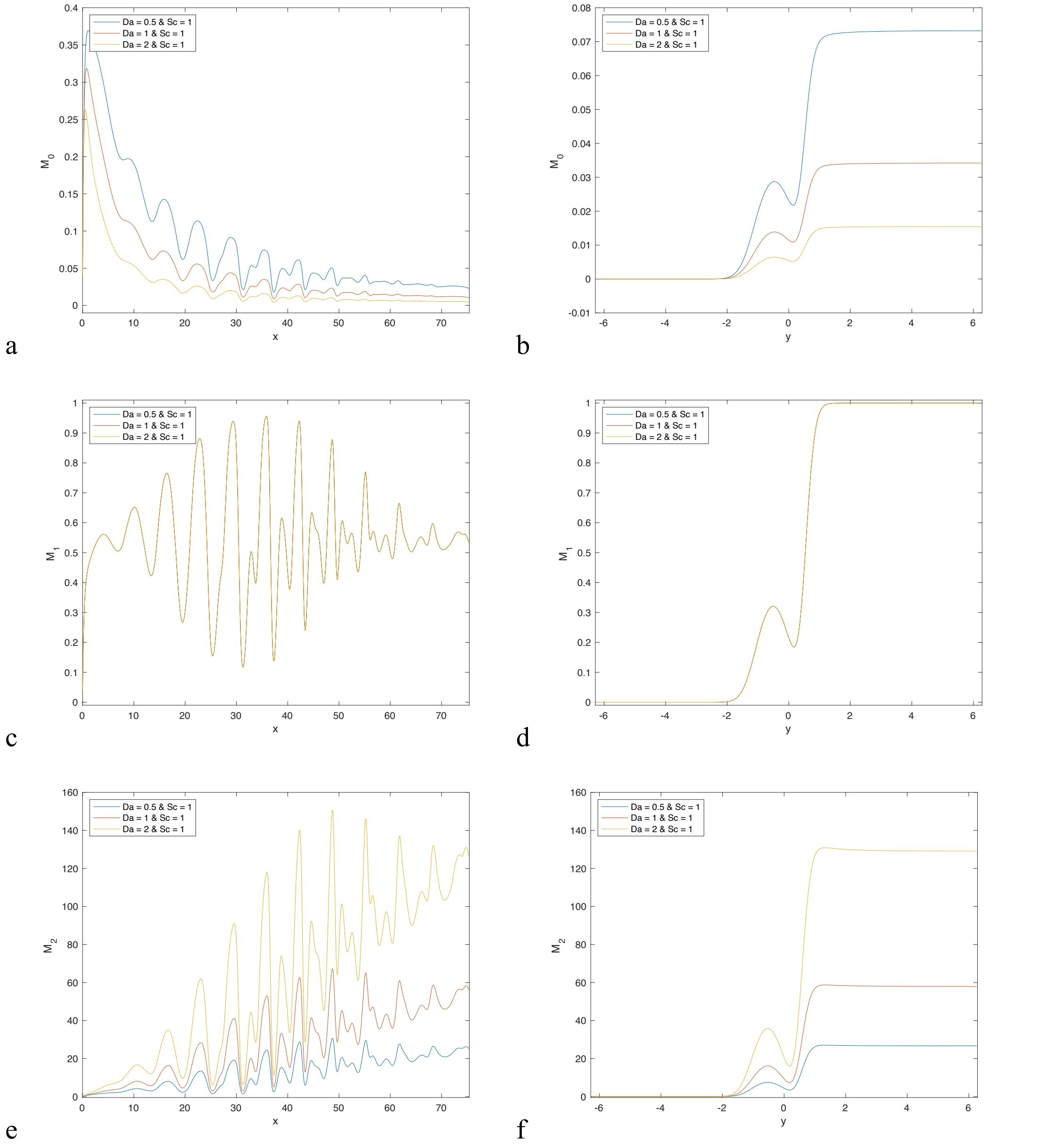

Figure 18. The effect of Damkohler number ($Da$) on the distribution of particle moment in space under the conditions $Re = 200$, $Sc = 1$. a) $M_0$ along the central horizontal line; b) $M_0$ along the central vertical line; c) $M_1$ along the central horizontal line; d) $M_1$ along the central vertical line; e) $M_2$ along the central horizontal line; f) $M_2$ along the central vertical line.

**Declaration of interest statement**

The authors report no conflict of interest.

Mingliang Xie
State Key Laboratory of Coal Combustion, Huazhong University of Science and Technology, Wuhan 430074, China
Correspondence Email: mlxie@mail.hust.edu.cn